\makeatletter 
\@namedef{ver@breakurl.sty}{}
\makeatother

\documentclass[]{jot}
\usepackage{multirow}

\jotdetails{
    volume=vv,      % volume
    number=nn,       % number or issue
    articleno=aa,   % article number, eg a1 for research articles, e for editorials
    year=2024,      % year
    license=ccbyncnd    % choose from ccby, ccbynd, ccbyncnd
}

\usepackage[english]{babel}
\usepackage{microtype} % optional, for aesthetics
\usepackage{tabularx} % nice to have
\usepackage{booktabs} % necessary for style

\newcommand{\e}[1]{\mbox{\lstinline[basicstyle=\normalsize,language=OOSC2Eiffel]|#1|}}

\usepackage{boogie}
\usepackage{eiffel}
\usepackage{listings}
\usepackage{enumitem}
\usepackage{tabto}
\usepackage{float}
\usepackage{graphicx}
\graphicspath{{./figures/}}
\usepackage{longtable}
\usepackage{tcolorbox}
\usepackage[framemethod=TikZ]{mdframed}
\usepackage{setspace}
\usepackage{xurl}
\mdfdefinestyle{MyFrame}{%
        linecolor=grey!50!white,
        outerlinewidth=2pt,
        roundcorner=20pt,        innertopmargin=6pt,%\baselineskip   
        innerbottommargin=6pt,%\baselineskip
        innerrightmargin=20pt,
        innerleftmargin=20pt,
        backgroundcolor=white}
    
\usepackage{color}      % for comments
\usepackage{amssymb}    % for comments
\newcommand{\mynote}[3][black]{\textcolor{#1}{\fbox{\bfseries\sffamily\scriptsize{#2}}
{\small$\blacktriangleright$\textsf{\emph{#3}}$\blacktriangleleft$}}}

\newcommand{\TODO}[1]{\mynote[red]{TODO}{#1}}
    \renewcommand{\TODO}[1]{} % to remove all the ToDos

\usepackage{flushend}

\usepackage{float}
\usepackage{flafter}
\usepackage{afterpage}

\usepackage{comment}
\usepackage{array}
\usepackage{amsthm}

\usepackage{lscape}
\usepackage{url}

\usepackage{xcolor}
\usepackage{caption}
\usepackage{subcaption}
\usepackage[english]{babel}
\usepackage[utf8]{inputenc}
\usepackage[T1]{fontenc}
\usepackage{stfloats} %for floating figures

\articletype{regular} 

\title{UOOR: Seamless and Traceable Requirements}

\author[$\ast$]{Maria Naumcheva}
\author[$\ast$]{Sophie Ebersold}
\author[$\ast$]{Jean-Michel Bruel}
\author[$\S$]{Bertrand Meyer}
\affil[$\ast$]{IRIT, University of Toulouse, France}
\affil[$\S$]{Constructor Institute of Technology, Schaffhausen, Switzerland and Constructor University Bremen, Germany}

\runningtitle{UOOR: Towards Seamless and Traceable Requirements} % For use in the internal pages 

\runningauthor{M. Naumcheva, S. Ebersold, J.-M. Bruel, B.
Meyer.}

\newenvironment{itemize*}%
  {
  \vspace{-6pt}
  \begin{itemize}%
    \setlength{\itemsep}{0pt}%
    \setlength{\parskip}{0pt}}%
  {\end{itemize}}

\usepackage[acronym]{glossaries}
\newacronym{nl}{NL}{Natural Language}
\newacronym{uoor}{UOOR}{Unified Object-Oriented Approach for Requirements}
\newacronym{uml}{UML}{Unified Modeling Language}
\newacronym{rsml}{RSML}{Requirements-Specific Modeling Language}
\newacronym{ireb}{IREB}{International Requirements Engineering Board}
\newacronym{babok}{BABOK}{Business Analysis Body of Knowledge}
\newacronym{srs}{SRS}{software Requirements Specification}
\newacronym{oo}{OO}{Object-Oriented}
\newacronym{sysml}{SysML}{Systems Modeling Language}
\newacronym{ears}{EARS}{Easy Approach to Requirements Syntax}
\newacronym{lms}{LMS}{Library Management System}
\newacronym{ocl}{OCL}{Object Constraint Language}
\newacronym{swebok}{SWEBOK}{Software Engineering Body of Knowledge}
\newacronym{rucm}{RUCM}{Restricted Use Case Modeling}
\newacronym{ucm}{UCM}{Use Case Maps}
\newacronym{asm}{ASM}{Abstract State Machines}
\newacronym{lotos}{LOTOS}{Language of Temporal Ordering Specification}
\newacronym{ooad}{OOAD}{Object-Oriented Analysis and Design}
\newacronym{gore}{GORE}{Goal-Oriented Requirements Engineering}
\newacronym{tdd}{TDD}{Test-Driven Software Development}
\newacronym{bdd}{BDD}{Behavior-Driven Development}
\newacronym{acl}{ACL}{Another Contract Language}
\newacronym{ide}{IDE}{Integrated Development Environment}
\newacronym{soor}{SOOR}{Seamless Object-Oriented Requirements}
\newacronym{soort}{SOORT}{Seamless Object-Oriented Requirement Templates }
\newacronym{eis}{EIS}{Eiffel Information System}
\newacronym{uri}{URI}{Uniform Resource Identifier}
\newacronym{gps}{GPS}{Global Positioning System}
\newacronym{imu}{IMU}{Inertial Measurement Unit}
\newacronym{gnss}{GNSS}{Global Navigation Satellite System}
\newacronym{ecu}{ECU}{Electronic Control Unit}
\newacronym{elem}{ELEM}{Project Element}
\newacronym{cstr}{CSTR}{Constraint}
\newacronym{fr}{FR}{Functional Requirement}
\newacronym{nlrq}{NLRQ}{Natural Language Requirement}
\newacronym{im}{IM}{Implementation Artifact}
\newacronym{sircod}{SIRCOD}{The Seamless Intergration of Requirements in Code}
\newacronym{doors}{DOORS}{Dynamic Object-Oriented Requirements System}
\newacronym{pegs}{PEGS}{Project, Environment, Goals, System}
\begin{abstract}
In industrial practice, requirements are an indispensable element of any serious software project. In the academic study of software engineering, requirements are one of the heavily researched subjects. And yet requirements engineering, as practiced in industry, makes shockingly sparse use of the concepts propounded in the requirements literature. The present paper starts from an assumption about the \textit{causes} for this situation and proposes a \textit{remedy} to redress it. The posited explanation is that \textit{change} is the major factor affecting the practical application of even the best-intentioned requirements techniques. No sooner has the ink dried on the specifications than the system environment and stakeholders' views of the system begin to evolve. %Requirement methods that assume that requirements can be set once and for all to guide the development are doomed.
The proposed solution is a requirements engineering \textit{method}, called UOOR, which unifies many known requirements concepts and a few new ones in a framework entirely devised to accommodate and support seamless change throughout the project lifecycle.

The Unified Object-Oriented Requirements (UOOR) method encompasses the commonly used requirements techniques, namely, scenarios, and integrates them into the seamless software development process. 
The work presented here introduces the notion of seamless requirements traceability, which relies on the propagation of traceability links, themselves based on formal properties of relations between project artifacts. As a proof of concept, the paper presents a traceability tool to be integrated into a general-purpose IDE that provides the ability to link requirements to other software project artifacts, display notifications of changes in requirements, and trace those changes to the related project elements. 

The UOOR approach is not just a theoretical proposal but has been designed for practical use and has been applied to a significant real-world case study—a Roborace that focuses on developing software for autonomous racing cars.

\end{abstract}

\begin{document}
\maketitle

\section{Introduction}

Software handles an increasing number of tasks, from enterprise management and reporting to operating medical devices. 
Before we can rely on software, however, we must be sure that it does exactly what it is supposed to do. 
The process of description of what the software will do and how it will perform is known as software requirements specification. 
This process plays a major role in the software development lifecycle: according to an often-cited study \cite{hussain2016role}, requirements-related factors are among the leading reasons for software project failures. 

An earlier article \cite{46} analyzed some limitations of dominant practices of requirements engineering, particularly use cases. More generally, challenges in the current state of the art (see section \ref{approaches} for further analysis) include the following: 
\begin{itemize}
    \item Requirements are \textbf{subject to change}. Requirements are rarely definitive at the beginning of the project; they, most of the time, evolve and change.
    \item The ability to adapt to changes in requirements is tightly related to the \textbf{traceability} between requirements and other project artifacts. Due to the necessity to manually create and maintain traceability links, requirements traceability is often perceived as an inefficient practice and is applied only in 40\% of software projects.
    \item Producing “good” requirements requires substantial effort. \textbf{Balancing requirements efforts} and the \textbf{quality} of the obtained requirements is difficult: quality requirements require certain rigor and formality, which requires substantial training for requirements engineers. Conversely, affordable approaches are much easier to grasp but leave much room for deficient requirements.
\end{itemize}

To help address these issues, we have developed an approach, Unified Object-Oriented approach to Requirements (UOOR), which
includes the following concepts:
\begin{itemize}

    \item \textbf{Object-oriented decomposition techniques}. We demonstrate, that the concept of class is general enough to describe important artifacts such as scenarios and  test cases. 
    \item \textbf{Formulating requirements with} \textbf{contracts}. Contract-based specification is implementation-independent and ensures unambiguity, verifiability and reusability of requirements.
    \item \textbf{Seamless software development}. Software development, from requirements to implementation relies on a uniform process (based on refinement) and a uniform notation (Eiffel language). 
    %, that can satisfy the need to quickly adapt to changes in requirements, ensuring fast and smooth traceability between requirements and other software artifacts.
    \item \textbf{Seamless requirements traceability}. The burden of traceability links creation and management can be lowered by the  propagation of traceability links, themselves based on formal properties of relations between project artifacts.
\end{itemize}

Section~\ref{characteristics} discusses the desired characteristics of a requirements approach.
Section~\ref{UOOR_chapter} introduces the UOOR approach and provides guidance on producing UOOR specifications. 
Section~\ref{toolchapter} explores the notion of seamless requirements traceability. 
It devises the typed relations between project elements and the formal properties of such relations. 
Section~\ref{traceability_tool} presents a tool to be integrated into EiffelStudio, which facilitates managing requirements traceability links. 
Sections~\ref{Roborace} and~\ref{Experiment} evaluate the UOOR approach. 
Section~\ref{Roborace} presents the application of the UOOR approach to a significant project -- the Roborace. 
Section~\ref{Experiment} reveals the results of a controlled experiment conducted at the University of Toulouse and evaluates the perception of the approach. 
Section~\ref{approaches} presents related work.
Section~\ref{contributions} summarizes and evaluates the contributions, lists the limitations of the devised approach, and highlights the perspectives for future work.

\section{Desired characteristics of a requirements approach}
\label{characteristics}

This section explores the characteristics of a requirements engineering approach necessary to make it usable in the industry.

Section~\ref{quality} explores some of the qualities of ``good'' requirements. Section~\ref{usable} investigates the characteristics of the requirements approach that make it usable.  

\subsection{Qualities of ``good'' requirements}
\label{quality}
According to the IEEE standard \cite{iso2018} requirements should be necessary, appropriate,  unambiguous, complete, singular, feasible, verifiable, correct, conforming, and traceable. Nevertheless, it is hard to achieve some of these characteristics if the requirements are documented in informal natural language. In contrast, according to the industrial survey conducted by Fricker et al. \cite{fricker2015requirements}, only 6\% of projects utilize formal notations. %In particular, \mb{we highlight such requirements properties as} verifiability, unambiguity and traceability. 
Besides, verifiability, unambiguity, and traceability are requirements properties that appear fundamental in a requirements analysis approach. 

\subsubsection{Verifiability} 
A requirement is verifiable if and only if there exists a process that can prove that the system satisfies the specified requirement \cite{ieee1998}. Although the notion of verifiability is defined in IEEE standard, not much guidance is provided on how to achieve this property apart from the statement that \textit{``verifiability is enhanced when the requirement is measurable''} \cite{iso2018}. 
% \mb{See where to refer to the verification survey -> added to section \ref{verifiability}}

\subsubsection{Unambiguity} 
Ambiguity refers to possible different interpretations of one and the same requirement. Natural language, which is a predominant way of specifying requirements, is an innate source of ambiguity \cite{34}. The only way to avoid ambiguity is to introduce some level of formality: to constrain the natural language or to utilize formal or semi-formal methods and notations. At the same time, formal methods are not widely used. The key reasons for that are that, on the one hand, formal methods require learning related notation and modeling techniques, which often require a strong mathematical background, and on the other hand, formal specifications are not readable, which makes it difficult to share the information and ensure specification accuracy. In view of this, we can clearly identify the need for a method to produce unambiguous requirements and specifications. 

\subsubsection{Traceability} 
Requirements traceability is the ability to follow both the sources and consequences of requirements in the rest of the product \cite{Handbook}. Requirements traceability ensures that the impact of changes in requirements specification is easily localizable in the code, which significantly decreases the time and costs of assessing the impact of changes and implementing the changes. Traceability relations can also be used to verify that the system meets its requirements. 

Even in regulated or safety-critical domains, such as healthcare or the military, traceability links are often created at the very end of the process and contain many deficiencies, such as incomplete, missing, or erroneous traceability data \cite{cleland2014software}. In less formal projects, traceability is often perceived as a ``made up problem'' or an ``unnecessary evil'' \cite{cleland2014software}. Consequently, only 40\% of software projects practice requirements traceability \cite{fricker2015requirements}. 

At the same time, the projects with missing traceability links are vulnerable to change and evolution: it can be tremendously hard to identify all the system elements where the changes should be introduced.

\subsection{Characteristics of a usable approach to requirements}
\label{usable}
A requirements approach is a strategy for developing and managing project requirements together with supporting guides and tools. To be used in industry, an approach must meet certain criteria, which will be explored in further subsections.

\subsubsection{Documentation}

To compete with established, extensively documented approaches to requirements, a new candidate must provide potential users with a clear, actionable description of how to apply the methodology \cite{methodology_role}.

%There exist many approaches to requirements; the most mature  ones are covered in textbooks, video tutorials, and university curricula. In order to switch from such an approach to a new one, a person would expect to find detailed guidance in the supporting materials on how to apply the approach. It was demonstrated that the application of a new requirements methodology is more efficient when requirements engineers use the procedural knowledge of an approach (i.e., the examples of the application of the methodology) \cite{methodology_role}.

\subsubsection{Ease of learning}
In many cases, the primary competitor to a proposed requirements approach is natural language, which does not require any learning. While any more formal method will require learning new concepts and notations, the effort should remain commensurate with the expected benefits \cite{davis2013study}. 
%Requirements specification approaches place different demands on the qualification of requirements engineers. For example, natural language requirements do not require specific knowledge, whereas formal methods require significant training. Too high prerequisites on requirements engineers’ background may hinder the method's adoption \cite{davis2013study}.

\subsubsection{Tool support}
Relying on tool-supported requirements engineering facilitates capturing, tracing, analyzing, and managing changes to requirements \cite{ibm_rq}. According to state of practice surveys \cite{kassab2015, RE_in_china, franch2023state}, 40 to 67\% of projects use requirements specification and management tools, with a tendency that the access to such tools is better in large multi-national corporations \cite{RE_in_china}. 

Tool support is essential for ensuring requirements traceability and managing changes in requirements: manual performance of such tasks is tedious and decreases the project's agility.

\subsubsection{Requirement artifacts reusability} 
\label{reuse}

Reusability is the degree to which an asset can be used in more than one system \cite{iso2017}. Reusability is widely adopted in software engineering: libraries of reusable components have become an integral part of programming languages. Requirements reuse has not reached a comparable level.

The study of Irshad et al. \cite{reusability_survey} reviews approaches to requirements reuse. The majority of the approaches (57\%) suggest reusing requirements in a textual form, or the form of requirements reuse is undefined. Other forms of requirements reuse include templates, use cases, modeling language-based artifacts, formal models, and features. In practice, requirements reuse is mostly limited to copying and modifying natural language requirements from previous projects \cite{palomares2017requirements}. %Improper reuse or requirements is dangerous: simple copy-pasting requirements from preceding projects may lead to catastrophes. as was the case in Therac-25 project \cite{baase2008ethical}.

\subsubsection{Seamlessness}
\label{seamlessness}
Software development may rely on several different notations, such as natural language, modeling language, formal language, and programming language. The process of switching from one notation to the other, when not seamless, is prone to errors. Seamless software development uses a uniform method and notation throughout all activities, such as problem modeling and analysis, design, implementation, and maintenance \cite{33}. Seamless software development facilitates traceability between requirements and other software artifacts.

\section{Unified Object-Oriented approach to Requirements}

\label{UOOR_chapter}

The idea of capturing software requirements with contracts has given  rise to several approaches \cite{32, 39, galinier2021seamless}. They address specific aspects of OO requirements and fall short of the goal of this article: offering a comprehensive requirements methodology.  
Starting from such ideas as Multirequirements \cite{32}, SOOR \cite{39}, SIRCOD \cite{galinier2021seamless} and PEGS \cite{Handbook}, the proposed methodology, Unified Object-Oriented approach to Requirements (UOOR) benefits from OO techniques, which have proven to be successful in software implementation, and scenarios,  commonly used for requirements. 

The  approach   unifies scenarios as a commonly used requirements technique with software contracts as a requirements formalization approach, which does not require a specific background in formal methods \cite{naumcheva2022object}. The reasons for combining these two techniques are the following:
\begin{itemize*}
    \item Scenarios are widely used in industry and are perceived as an excellent tool for eliciting requirements and communicating with project stakeholders.
    \item Scenarios, formulated in natural language, are ambiguous and lack abstraction.
    \item Software contracts provide means for requirements formalization.
    \item Software contracts do not rely on a specific mathematical notation and are as easy to formulate as ``if ... then ...'' instructions. 
    Capturing requirements with contracts makes them verifiable, reusable and traceable.
\end{itemize*}

%As discussed earlier in section~\ref{seamlessness}, 
In line with the seamlessness principle,  we rely, for the formal requirements notation, on a statically typed object-oriented programming language: Eiffel   \cite{Eiffel}, chosen for its readability and support for contracts. Examples in this article will be expressed in Eiffel and  the tool support will be based on the facilities of the EiffelStudio IDE  \cite{19}. Any other statically typed OO language supporting contracts could be used; for example, the RQCODE approach provides a framework for seamless expression of security requirements in Java \cite{ildar2, ildar3}. 

The UOOR approach builds on the general idea of object-oriented requirements and adds its own specifics. The following concepts are common to all OO requirements approaches:

\begin{itemize*}
    \item Object types, described through applicable operations -- queries (providing information) and commands (updating information).
    \item Software contracts, which capture the semantics of operations.
\end{itemize*}

To these, UOOR adds:
\begin{itemize*}
    \item Specification drivers \cite{38}, which capture the system's behaviors (section~\ref{subsection4}).
    \item Support for  traceability through seamlessness (section~\ref{toolchapter}).
\end{itemize*}
UOOR describes requirements specification across three dimensions: object model; functional specification; behavioral specification.  

\begin{comment}
    (see Fig.~\ref{UOOR_model}).

\vspace{-6pt}
\begin{figure}[htb!]
    \centering
    \includegraphics[width=0.6\linewidth]{UOOR_model.jpg}
    \caption{Three dimensions of UOOR requirements.}
    \label{UOOR_model}
\end{figure}

\end{comment}

An \textbf{object model} captures key abstractions in the application domain in the form of software classes. This static model specifies abstract data types for all relevant environmental phenomena.

A \textbf{functional specification} provides a specification of individual operations. Operations correspond to features of classes of the object model. Their abstract specification relies on in-class contracts.

A \textbf{behavioral specification} defines permissible sequences of operations. Behavioral specification relies on the following mechanisms:

\begin{itemize*}
    \item Specification drivers capture scenarios as example sequences of operations and may serve testing purposes.
    \item Software contracts provide more abstract specifications of properties that would otherwise be expressed as time-ordering constraints.
\end{itemize*}

\begin{figure*}[!ht]
    \centering
    \includegraphics[width=\textwidth]{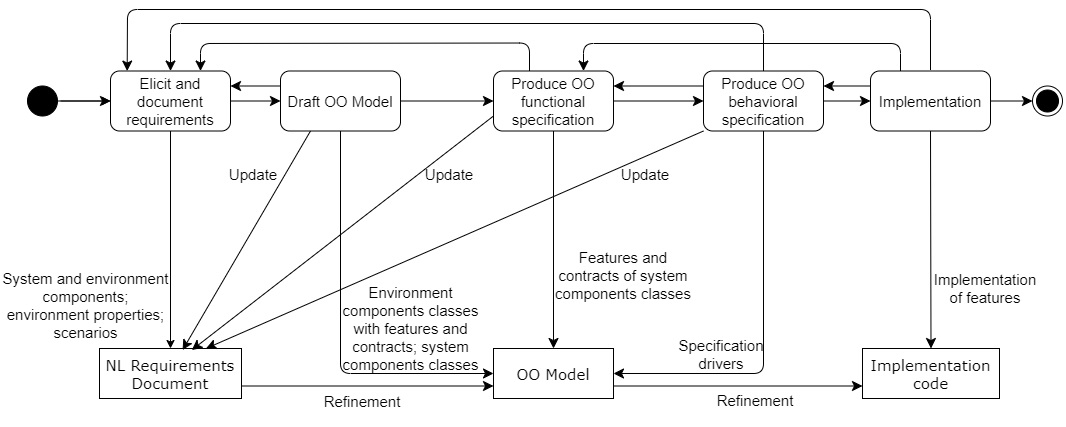}
    \caption{Overview of the UOOR approach.}
    \label{UOOR_overview}
\end{figure*}

Object-oriented principles pervade the whole approach --- not as an afterthought (as in ``let's make these requirements OO now!''), but as a fundamental modeling discipline applied from beginning to end.
%In other words, the idea is not to produce some traditional kind of requirements and then make the result object-oriented; instead, OO principles help structure the entire effort by providing a unifying conceptual framework for describing the system and its environment.

The overall process iterates the following  basic steps, each with a specified resulting product:
\begin{enumerate}
    \item Eliciting and documenting requirements, resulting in natural-language requirements document. (section~\ref{subsection1}).
    \item Devising a OO model, which captures the key components of the system and its environment (section~\ref{model_environment}).
    \item Devising functional OO requirements, which enrich the OO model with features expressing functional requirements and their contracts, as well as  environment properties and their constraints (section~\ref{subsection3}). 
    \item Behavioral specification, which adds abstract properties of the system's functions (such as time-ordering constraints) extracted from scenarios, as well as concrete scenario examples expressed as specification drivers (contracted routines of scenario classes), also useful for verification purposes (section~\ref{subsection4}). 
    \item Refinement, leading to implementation classes, typically inheriting  from requirements classes and hence satisfying the corresponding requirements-level contracts (section~\ref{subsection_refinement}). 
\end{enumerate}

Fig.~\ref{UOOR_overview} shows the overall UOOR process. The following sections~\ref{subsection1}~-~\ref{subsection4} provide the details of each step.

\subsection{Eliciting and documenting requirements}
\label{subsection1}
The UOOR approach does not advocate any particular methodology for eliciting and documenting requirements. One possible starting point is the PEGS template \cite{Handbook}, which organizes requirements into four books: Project (focusing on project management aspects and not essential for the present discussion), Environment, Goals and System. It has room for all important requirements elements, for example:

\begin{itemize*}
    \item System and environment components (in the System book).
        \item System functions (also in the System book).
    \item Environment assumptions, constraints, and invariants (in the Environment book).
    \item Scenarios (in the Systems or Goals book depending on the nature of the scenarios)..
\end{itemize*}
%Whereas any other template may be used to document requirements in the UOOR approach, it is important that it includes the aforementioned elements.
Any template, whether PEGS or another one, should document each requirement in a separate paragraph.
%Otherwise, the approach leaves freedom in the process of requirements elicitation and provides specific guidance on how to proceed with the elicited data (i.e., how to analyze scenarios and environment properties).

%The approach will also tackle the important task of tracing requirements from requirements documents to other project artifacts and in the reverse direction (see sections~\ref{subsection3},~\ref{traceability}).

\subsection{Modeling components of the system and its environment}
\label{model_environment}

To produce an OO model, a requirements engineer should describe key abstractions in the application domain with classes. These classes should cover both the system and its environment. We will use the example of a simple Library Management System (LMS) to illustrate the approach.

Examples classes are: \e{LIBRARY} (system), \e{BOOK} and \e{PATRON} (both environment).  During its operation, the system manipulates objects - instances of these classes. In the LMS, such objects will be computer representations of such library objects --- physical or conceptual --- as books, patrons and catalog. 

The PEGS template includes sections for environment components (E.2 in the Environment book) and system components (S.1); for traceability, we can link their natural-language components with elements of the corresponding classes in the OO model (details appear in below in sections~\ref{traceability} and ~\ref{traceability_tool}).

Object technology supports modeling systems at various levels of abstraction. In particular, both a feature and a class may be either effective or deferred\footnote{The respective terms ``concrete'' and ``abstract'' are also used.}. ``Effective'' means fully implemented (and hence ready to execute as part of a program). A feature or class is “deferred” if its definition does not include  an implementation or (for a class) includes a partial implementation only; it may still, however, include an abstract specification of the properties of the feature or class, in the form of a ``contract'' as explained below. Deferred features and classes are particularly useful for requirements since requirements focus on specifying behavior rather than implementing it. For example, \e{LIBRARY_ITEM} may be a deferred class since it describes an abstract concept with several possible concrete realizations, such as \e{BOOK}. In some cases, requirements classes have to be effective, in particular, scenario classes and test classes.

Client and inheritance relations capture the relations between the environment and system components. For example, operations on a book  involve, among other objects, a patron and a library. The class representing a book will be a client of the
classes representing the other two concepts. 
A class inherits from another if it represents a specialized
or extended version of the other's concept. For example, books
and magazines both belong in the general category of library
items, which can be represented by a class \e{LIBRARY_ITEM}.
\e{BOOK} and \e{MAGAZINE} are descendants of \e{LIBRARY_ITEM}
through the inheritance relation. As another example, there could be two categories of patrons, regular and research patrons, with different restrictions such as the number of books on hold or the hold's duration. In that case, classes \e{REGULAR_PATRON} and \e{RESEARCH_PATRON} would
inherit from the class \e{PATRON}, which captures features and contracts applicable to all patrons.

Each class defines the properties of the corresponding objects through the operations applicable to them: queries and commands. Queries provide information about objects, whereas commands update the corresponding objects. 

A specification of the properties of systems and their objects,  through the list of associated classes and their features, only gives structural properties. To provide the actual semantics of these elements — other than through the implementation – we should also express their abstract properties. Contracts fulfill this need. They include:

\begin{itemize*}
    \item Preconditions of a feature, specifying the conditions under which it can be used.
    \item Postconditions of a feature, expressing properties resulting from its application.
    \item Class invariants, expressing consistency properties applicable to all instances of a class.
\end{itemize*}

Such a contract element consists of assertions, each of which is an individual boolean property applicable to the corresponding objects, such as (for a \e{BOOK} instance) \textit{``the book is currently on loan''}. We rely on contracts to express:
\begin{itemize*}
    \item Properties of the environment, such as \textit{``a patron can reserve no more than 5 books”} (discussed below in the present section). 
    \item Properties of the system's functions, such as \textit{``if a patron has less than five holds, placing a hold is successful and the
book becomes reserved''} (discussed in section 
\ref{subsection3}).
\end{itemize*}

The important part of the OO model is the properties of the system's environment. According to \cite{Handbook} those properties are of  three kinds:
\begin{itemize*}
    \item Assumptions - properties, that are satisfied by the environment, so the system takes them for granted (\textit{``The length of a book title does not exceed 100 characters''}).
    \item Constraints - obligations coming from the environment that the system must comply with (\textit{``a patron can reserve no more than 5 books''}).
    \item Invariants - environment properties that the system must maintain (\textit{ ``A book can have exactly one of the following statuses: available, on hold, borrowed, due''}).
\end{itemize*}

Various types of environment properties play different roles in software development. 

Assumptions facilitate the work of the developers,  by restricting the set of cases to be handled by the system, as in  \textit{``The length of a book title does not exceed 100 characters''}. They are engineering decisions and may need re-examination in later stages of the project; stating them explicitly is critical to assess their consequences and uncover possible violations, particularly at the time of system validation and verification. (The famous Ariane-5 software failure was due to an unspotted obsolete assumption.) 

%provides information on the size of manipulated data when performing operations with the book titles. At the same time, assumptions (and this one in particular) are falsifiable: they can be wrong. Expressing the assumptions explicitly with assertions helps in verifying their validity: if during system verification and validation, the assertion expressing the system assumption is violated, the respective error message will be displayed. The developers will then be able to assess and address the consequence of reconsidering the assumption. 

Constraints and invariants express the properties of the environment that the system must respect. Such constraints can be expressed in the associated class texts as class invariants. The  constraint \textit{``a patron can reserve no more than 5 books''} can be expressed as the following invariant clause in class \e{PATRON}: \textit{num\_reserved <= 5 }. Listing~\ref{book_class} provides another example.

\begin{comment}
\begin{lstlisting}[caption={Implementation of an invariant \textit{``A patron can reserve no more than 5 books''.}}, captionpos=b, label=5books]
deferred class 
    PATRON
feature
    num_reserved: INTEGER
       -- Number of books reseved by the patron
invariant
    num_reserved <= 5
end
\end{lstlisting}
\end{comment}

\begin{lstlisting}[caption={Expressing the invariant property \textit{``a book can have exactly one of the following statuses: available, on hold, borrowed, due.''}}, captionpos=b, label=book_class,language=OOSC2Eiffel]
deferred class BOOK feature
    is_available, is_on_hold, is_borrowed, is_due: BOOLEAN
invariant
    is_available implies not (is_on_hold or is_borrowed or is_due)
    is_on_hold implies not (is_available or is_borrowed or is_due)
    is_borrowed implies not (is_available or is_on_hold or is_due)
    is_due implies not (is_available or is_on_hold or is_borrowed)
end
\end{lstlisting}

%(listing~\ref{5books}):

\vspace{12pt}
We can use pre- and postconditions to express properties of operations on the components of the environmen, such as, in class \e{PATRON}, \textit{``a patron can place hold on a book only if the book is available.''}

This constraint is implemented as a precondition of the feature \e{place_hold} of the class \e{PATRON} (listing~\ref{precond_constraint}):

\begin{lstlisting}[caption={Implementation of the constraint \textit{``A patron can place hold on a book only if the book is available.''}}, captionpos=b, label=precond_constraint,language=OOSC2Eiffel]
deferred class PATRON feature
    num_reserved: INTEGER
        -- Number of books reserved by the patron       
    place_hold(b: BOOK)
        require 
            b.is_available
        deferred
        end
invariant 
    num_reserved <= 5
end
\end{lstlisting}

%The constraint, related to the feature ``place hold'' saying that a patron cannot reserve more than 5 books, is implicit. But what happens if he or she has 5 books reserved and attempts to reserve one more? One option is that the book reservation will be denied. Another option is to cancel the oldest hold and complete the last one. To avoid the ambiguity we add one more constraint (which is in fact an implicit business rule): \textit{``A patron can place hold on a book if she has less than 5 books on hold.''}

%To express this constraint, we add one more precondition to the feature \e{place_hold} %(listing~\ref{patron_precond}):

\begin{comment}
    
\begin{lstlisting}[caption={Implementation of the constraint \textit{``A patron can place hold on a bookif she has less than 5 books on hold.''}}, captionpos=b, label=patron_precond]
deferred class 
    PATRON
feature
    num_reserved: INTEGER
       -- Number of books reserved by the patron
       
    place_hold(b: BOOK)
        require
            b.is_available
            num_reserved < 5
        deferred
        end
invariant
    num_reserved <= 5
end
\end{lstlisting}
\end{comment}

\subsection{Producing an OO functional specification}
\label{subsection3}

In object-oriented requirements, elements of system functionality are expressed by features of classes. Contracts capture theif properties.

Consider the following requirement: \textit{``The Library Management System shall provide the ability to place a hold on books''}. To express it,
%,requirement%in an object-oriented style
we add a feature \e{place_book_on_hold} to the \e{LIBRARY} class (listing~\ref{library_precond}).

\begin{lstlisting}[caption={Implementation of the requirement \textit{``The Library Management System shall provide the ability to place hold on books.'' }}, captionpos=b, label=library_precond,language=OOSC2Eiffel]
deferred class 
    LIBRARY
feature       
    place_book_on_hold (b, p)
      -- Reserve a book b by patron p
        deferred
        end

\end{lstlisting}

Other system specifications follow from environment constraint, such as: \textit{A patron is limited to five holds at any given moment}. This constraint, already expressed in the OO model through the invariant of the environment class \e{PATRON}, yields a system property, governing the operation \e{place_book_on_hold}:
\begin{itemize*}
    \item If hold is allowed, placing a hold is successful and the book becomes reserved.
    \item If hold is not allowed, placing a hold is not successful, and the book remains available.
\end{itemize*}

Listing~\ref{place_hold_constraints} provides the OO implementation of these two constraints.

\vspace{-6pt}
\begin{lstlisting}[caption={Implementation of the constraints on the \e{place_book_on_hold} feature.}, captionpos=b, label=place_hold_constraints,language=OOSC2Eiffel]
    
deferred class LIBRARY feeature
   place_book_on_hold (b: BOOK; p:PATRON)
    -- Reserve a book b by patron p        
      deferred
      ensure
         old p.num_reserved < 5 implies 
           (b.is_on_hold and 
            b.patron.is_equal(p) and 
            p.num_reserved=old p.num_reserved+1)
         old p.num_reserved >= 5 implies 
           (b.is_available and 
            p.num_reserved = old p.num_reserved)            
      end

\end{lstlisting}

We should also link object-oriented requirements, expressed in an OO model, to  their counterparts in a natural-language requirement document. Hyperlinks between the two kinds of documents, discussed in section~\ref{traceability_tool}, will fulfill that purpose. 
%Clicking the hyperlink from the class code will open the requirements document at the bookmark, corresponding to the respective requirement. Clicking the hyperlink in the requirements document will open the source code at the feature that is specified by the given requirement. 

\subsection{Producing an OO behavioral specification} 
\label{subsection4}

%As noted in section~\ref{approaches}, it is common to rely on use cases or other forms of scenarios to model system behaviors. These techniques are not, however, the only way to describe behavior. In this section, we review how UOOR helps achieve this goal. 

The behavioral specification in UOOR relies on:
\begin{itemize}
    \item Expressing concrete scenario examples as specification drivers.
    \item Extracting abstract properties of operations from scenarios (such as time-ordering constraints).
\end{itemize}

A scenario, also known as a use case, is a pattern exercising the features (operations) of one or more classes. Use cases are procedural rather than object-oriented, but complement the OO model by describing practical examples of use of its abstractions and are particularly useful for both requirements elicitation and for system testing. In the  Use Case 2.0 approach\cite{26}, a successor to Ivar Jakobson's original use case, a \textit{use-case story} describes a possible path through a use case that is of value to a user or other stakeholder. 

UOOR expresses such a story as a routine, exercising the features of the target classes. The routines, specifying a set of related use case stories can be grouped in a separate class, providing ``specification drivers'' \cite{39}.
%, presented in section~\ref{contract_based}.
The following is an example, with specification drivers exercising features of classes \e{BOOK}, \e{PATRON} and \e{LIBRARY} (listing~\ref{use_case_stories_class}):

\begin{lstlisting}[caption={Use case stories class.}, captionpos=b, label=use_case_stories_class, language=OOSC2Eiffel]
class LIBRARY_BOOK_USAGE_STORIES feature  
    reserve_book_successfully (b: BOOK; lb: LIBRARY; p: PATRON) 
        require
            p.num_reserved < 5
            b.is_available
        do
            lb.place_book_on_hold (b, p) 
        ensure
            b.is_reserved 
            b.patron = p
        end             
    reserve_book_num_holds_exceeded (b: BOOK; lb: LIBRARY; p: PATRON)
        --See the implementation in a Github repo       
    -- Other use case stories
end

\end{lstlisting}

Expressed in this OO style, use case stories double down as test cases when provided with actual values for their arguments. If we call \e{reserve_book_successfully}, using actual instances of \e{BOOK}, \e{LIBRARY} and \e{PATRON} as arguments, we get a test case for that story. The arguments must satisfy the preconditions; for the routine to be correct, execution must satisfy the postcondition on exit.

OO techniques avoid \textit{premature} \textit{time-ordering decisions}. While it is possible for an OO specification to state a time-ordering constraint through a scenario, object technology also supports a more general and more abstract specification style based on contracts.
%Scenarios specify the order in which operations will be executed. Enforcing such an ordering specification at the level of requirements is often a premature decision. In reality, the order of the steps is not cast in stone. Using a preset ordering is convenient for describing desirable scenarios or, more generally, the expected ones. But what happens in life is not always what we hope for or expect. What if the customer returns a damaged book?
%Should the book not remain unavailable until it is repaired? Extensions can be used to specify scenarios that depart from the standard ones. However, this solution does not scale. Writing ever more use case extensions to cover all such situations leads to an explosion of special cases which soon become intractable. In practice, it is possible to write use cases to cover the most common scenarios, but they are only a small subset of the possible ones, in the same way that, in programming, tests can only cover a minute subset of possible inputs.
The idea is that instead of a use case strictly specifying the sole possible order $o_1$, $o_2$, $...$, we specify assertions $p_i$ and $q_i$ serving respectively as precondition and postcondition of $o_i$. To specify the indicated exact ordering, we just take $p_{i+1}$ to be the same as $q_i$; but by playing with the $p_i$ and $q_i$ we can much more flexibly specify a wide range of ordering constraints rather than prematurely prescribing just one obligatory order. 

%Class \e{BOOK} specifies these logical constraints in the form of contracts (listing~\ref{time_ordering_constraints}).  

\begin{lstlisting}[caption={Main scenario.}, captionpos=b, label=main_scenario, language=OOSC2Eiffel]
-- Main scenario of the use case ``borrow a book'':
    place_hold (patron: PATRON)
    checkout (patron: PATRON)
    return (patron: PATRON)

\end{lstlisting}

As an example, the specific sequence of actions described in the “Main scenario” use case (listing~\ref{main_scenario}) is compatible with the logical constraints specified in the class \e{BOOK}(list. \ref{time_ordering_constraints}), as postcondition of each step other than the first implies the precondition of the next one. It is an overspecification, however, prohibiting for example the inclusion of additional operations between \e{place_hold} and \e{check_out}. The logical properties expressed by the assertions of class \e{BOOK} relax this over-constraining order while preserving the necessary ordering constraints.

\begin{lstlisting}[caption={Illustration of logical constraints.}, captionpos=b, label=time_ordering_constraints, language=OOSC2Eiffel]
deferred class BOOK feature
    is_available, is_on_hold, is_checked_out, is_due: BOOLEAN     
    place_hold (p: PATRON)
	    -- Place a hold on a book. Set is_on_hold 
        require is_available 
        deferred 
        ensure
            is_on_hold 
            not is_available 
        end        
    checkout (p: PATRON)
        -- Check out the book 
        require is_on_hold 
        deferred 
        ensure is_checked_out 
        end        
    return 
        -- Return the book to the library 
        require is_checked_out or is_due 
        deferred 
        ensure is_available
        end 
end

\end{lstlisting}

\subsection{From requirements to code}
\label{subsection_refinement}

The seamless software development process works by iteratively refining requirements into executable code. Section~\ref{refinement} describes the details of that refinement process; ~\ref{traceability} explores how it supports requirements traceability.

\subsubsection{Refinement}
\label{refinement}
In the UOOR approach, the development process consists of refinement steps: elicited requirements are refined into OO requirements, which are further refined to implementation code. The steps are the following:
\begin{itemize*}
    \item Refine natural-language component requirements into deferred Eiffel classes, which will constitute the object model of the system and its environment. %The hyperlink in a requirements document links a component requirement with the respective class. The EIS note links the class  with a component requirement in a requirements document.\se{paragraph to be discussed- EIS link not defined before, Object model?, the hyperlink? ... In fact, No reason to speak from traceability there.}
    A hyperlink in the requirements document links the considered component requirement with its respective class
    \item Formulate environment constraints as contracts in classes, implementing the environment components. %The EIS note links environment constraint with the respective feature in the object model. 
    A note (using the Eiffel ``note'' construct for adding structured annotations to the code) links an environment constraint with a feature in the implementation (class model).
    \item Refine constraints further into functional requirements and constraints, expressed both in the natural-language document and (in the respective form of features and contracts) to the OO model.
    \item Proceed to implementation by writing effective classes, which in many cases will inherit from (deferred) requirements classes, providing implementations of their deferred features. The language's rules on inheritance (invariant conservation, precondition conservation or weakening, postcondition conservation or strengthening) guarantee that they satisfy the contracts formulated in the ancestor requirements classes.
\end{itemize*}

\subsubsection{Traceability links}
\label{traceability}
To ensure two-way traceability, a requirements engineer should create traceability links of two types: from natural-language (NL) documents to code artifacts, and the other way around.

Links of the first kind will appear in the natural-language documents as hyperlinks, pointing:
\begin{itemize*}
    \item From an environment or system component to its counterpart (a class) in the OO model.
    \item From a functional requirement to its counterpart (a feature with its contract) in the OO model.
    \item From another environment property, such as an assumption or constraint, to its counterpart (such as a class invariant) in the OO model.
\end{itemize*}

Links in the reverse direction point from a software artifact, such as a class or one of its features, to some part of a natural-language document. A traceability tool developed for the UOOR approach supports such links in the form of ``note'' code annotations; see its presentation in section~\ref{traceability_tool} below.

\subsection{System verification and requirements reuse}
\label{section_verification-reuse}
The UOOR approach integrates requirements with the rest of the entire software development process, providing the means for verifying the solution against the requirements (section~\ref{verifiability}), and supports reuse of domain-specific component requirements (section~\ref{reusability}).

\subsubsection{System verification}
\label{verifiability}
UOOR requirements enable both static and dynamic verification of the implemented system.

Since OO requirements are code elements, an IDE provides basic consistency checks at compile time. When contract checking is enabled at runtime, the IDE monitors contract violations. Since every contract can have a unique tag, a developer can trace an exception to the violated contract.

Contract specifications serve as oracles for dynamic testing. In addition, scenarios, implemented as specification drivers, serve as tests when passed actual arguments.

\begin{figure}[htb!]
    \centering
    \includegraphics[width=1.06\linewidth]{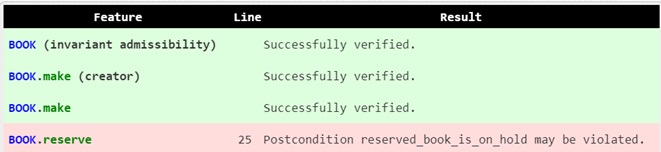}
    \caption{Autoproof output for the incorrect implementation of the class \e{BOOK}.}
    \label{autoproof}
\end{figure}

A static verifier, such as Autoproof \cite{7}, can be used to ensure static verification of the system's functional correctness, which significatly reduces testing costs \cite{huang2023lessons}.
\textbf{Verifiability} of a requirement refers to the ease of checking that the constructed system meets the requirement. Since in UOOR requirements are captured by contracts, EiffelStudio provides the following mechanisms for requirements verification:

\begin{itemize*}
    \item When contract checking is enabled at runtime, EiffelStudio monitors contract violations. Since every assertion can have a unique tag, a developer can trace an exception to the violated assertion.
    \item Contracts serve as test oracles, which simplifies producing tests. 
    \item Since requirements are compilable software elements, the IDE provides basic consistency checks such as type checking.
    \item Autoproof \cite{7} provides static verification facilities for contracted Eiffel programs. This tool checks whether the implementation satisfies its contracts. If some of the contracts may be violated, Autoproof outputs a warning, pointing at the contract that may be violated (see Fig.~\ref{autoproof}). 
\end{itemize*}

\vspace{-6pt}

\subsubsection{Requirement artifacts reusability}
\label{reusability}

Based on the capabilities provided by object-oriented technology, requirements obtained using the UOOR approach can be reused not by copy-pasting natural language texts but as libraries of domain-specific component specifications in the form of contracted deferred classes. Being implementation-independent, such specifications support different implementations.

UOOR organizes requirements around classes, which are abstractions of the objects in the application domain. Such abstractions can be organized into libraries and further shared between several projects. Thanks to genericity, requirements can be abstracted to generic modules, whose parameters represent types. The example of such a generic module is \e{CATALOG [CATALOG_ITEM]}, abstracted from a \e{BOOK_CATALOG} of the library case study, with such operations as ``has item?'', ``add item'', ``remove item'' (listing~\ref{reusable_class}). 

\begin{lstlisting}[caption={Example of a reusable requirements class.}, captionpos=b, label=reusable_class,language=OOSC2Eiffel]
deferred class CATALOG [CATALOG_ITEM] feature
    count: INTEGER
      -- Number of elements in catalog
    is_empty: BOOLEAN
      -- Is catalog empty?
        deferred
        ensure Result = (count = 0)
        end
    has (el: CATALOG_ITEM): BOOLEAN
      -- Is el an element of catalog?
        deferred
        ensure not_found_in_empty: Result implies not is_empty
        end
    add (element: CATALOG_ITEM)
      -- Add a new element to catalog
        deferred
        ensure count = 1 + old count
        end
    remove (element: CATALOG_ITEM)
      -- Remove element from catalog
        deferred
        ensure count = old count - 1
        end
end
\end{lstlisting}

Such a class defines basic operations applicable to objects of type catalog but stays free from implementation details. Due to its genericity, the class can be used for catalogs of different object types, not necessarily books or library items.

\section{Seamless requirements traceability}
\label{SeamlessTraceability}

\label{toolchapter}

\subsection{Seamless requirements traceability}

Requirements traceability is the ability  to follow both the sources and consequences of requirements \cite{Handbook}. Traceability involves \textit{traceability links}, connecting requirements with their sources and with related project elements.

%In practice, the navigation between the project elements, connected with a traceability link, can have a different nature. 

A basic technique consists of manually creating traceability matrices, by recording into a table individual links between requirements and test cases.
%Such matrices are created and updated manually, and requirements and test cases are added by copy-pasting from related documents.
Producing and maintaining such matrices requires substantial effort. They do not provide direct navigation from a requirement to a test case, but only establish a visual representation of the correspondence between requirements and related test cases. In their typical use they trace requirements not to general project artifacts but only to test cases. 

Dedicated requirements management tools (such as Polarion \cite{Polarion} and IBM DOORS \cite{69}) significantly decrease the requirements traceability effort. They allow the creation of clickable links between requirements and various project artifacts, for example by drag-and-drop. 
%In some cases it requires integration between several applications since project elements are not handled in a single application. 

\textit{Seamless requirements traceability} goes further. In addition to supporting direct links between requirements and other project artifacts, it relies on relation propagation, which is based on formal properties of relations. Three fundamental properties apply:
\begin{itemize*}
    \item Requirements are directly connected with other project artifacts with clickable links.
    \item A substantial number of links are propagated rather than created manually.
    \item Changes in requirements can be traced to the related project artifacts.
\end{itemize*}
%\cite{meyer1985software}

\subsection{UOOR traceability information model}

\begin{figure}[h!]
    \centering
    \includegraphics[width=\linewidth]{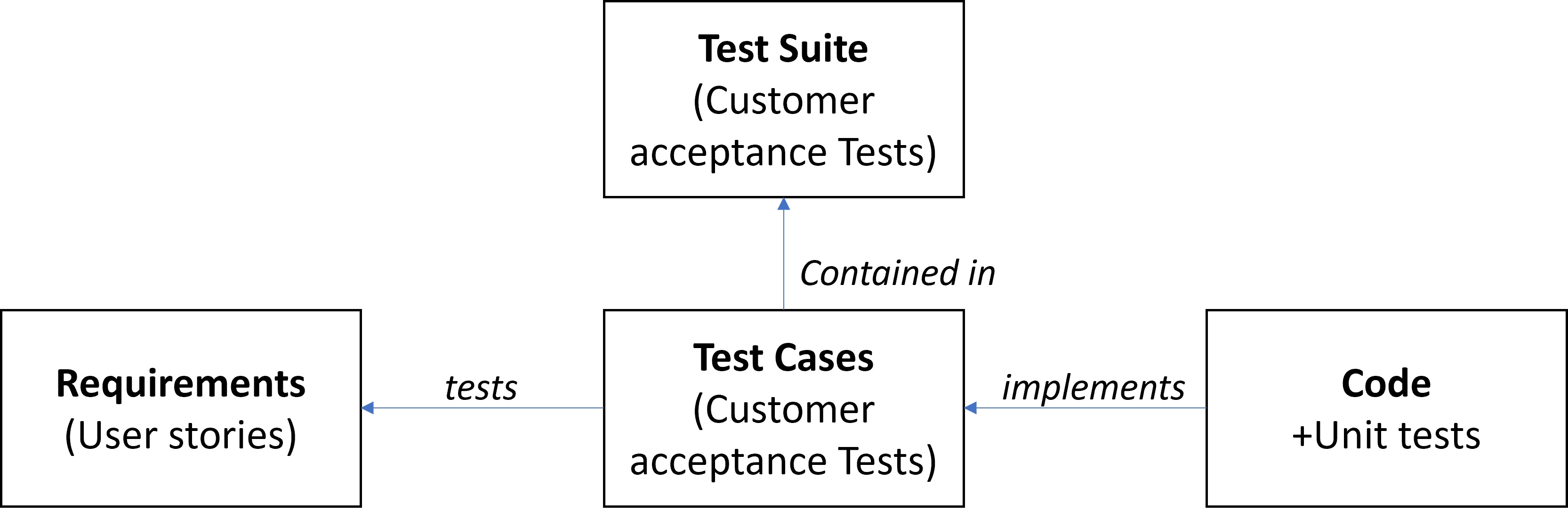}
    \caption{Agile traceability information model.}
    \label{agile_traceability_model}
\end{figure}

The traceability information model describes the relation between software project artifacts \cite{bunder2017model}. The basic traceability model has been established by the members of the Agile Project Management Forum \cite{cleland2011traceability} (see Fig.~\ref{agile_traceability_model}). This model reflects the common practice of agile projects by covering the most frequent tracing scenarios. 

The agile traceability information model includes three main blocks: requirements, tests (part of a test suite) and implementation. The relationship between requirements (user stories) and implementation is indirect: requirements are linked with test cases by external tools or by annotating test cases with user story ID. Between code and tests (and further to requirements), the relationship is implicit: we know that code implements the requirements if it passes the tests. 

\begin{figure}[h!t]
    \centering
    \includegraphics[width=1.1\linewidth]{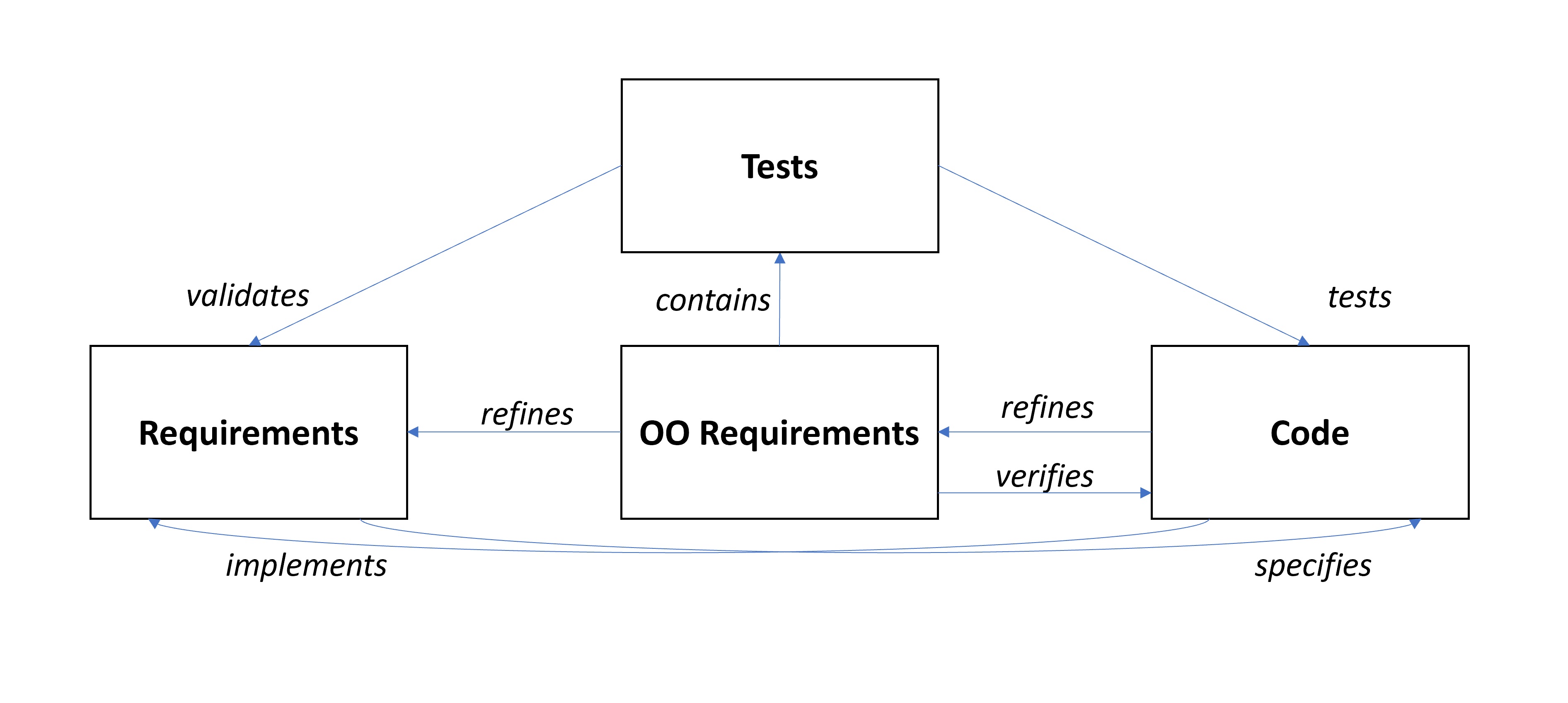}
    \caption{UOOR traceability information model.}
    \label{UOOR_traceability_model}
\end{figure}

Seamless development can extend this model by adding OO requirements, which link natural language requirements to tests and implementation code. In a coarse-grained view, the model looks as in Fig.~\ref{UOOR_traceability_model}.

Requirements in the UOOR traceability information model  are linked directly to the implementation  and test artifacts. To provide a more fine-grained overview of project elements and relations between them, sections~\ref{project_elements}-\ref{relations} define the main types of project elements and relations between them. 

\subsection{Project elements}
\label{project_elements}

Project elements are the basic blocks of the model of requirements engineering activity. They are artifacts produced in the course of a software development project. It is important to identify and define project element types for the following reasons:
\begin{itemize*}
    \item To guide their possible implementations (for example, as a class or a section of a document).
    \item To make sure that requirements can be traced to certain project artifacts (such as implementation or test artifacts).
    \item To introduce type-dependent propagation rules.
\end{itemize*}
As presented in Fig.~\ref{UOOR_traceability_model}, project artifacts belong to one of four general types: natural language requirements, OO requirements, tests and implementation. Table~\ref{project_element_types} provides the definitions of these four types and of project elements belonging to each of these groups. 

\begin{table*}[htbp]
\centering
\begin{tabular}{|p{0.2\textwidth}|p{0.46\textwidth}|p{0.25\textwidth}|}
\hline
\textbf{Project element type} & \textbf{Definition}                                                                                                      & \textbf{Possible implementation}                          \\ \hline
NL requirement         & \multicolumn{2}{|l|} {A requirement expressed in a textual form}%                                            & Bookmarked   section of a document                          
\\ \hline

Component requirement          & A statement   that the system or its environment contains a certain part                                            & Bookmarked   section of a document                          \\ \hline
Functional requirement       & A statement   that the system shall provide a certain functionality                                                      & Bookmarked   section of a document                          \\ \hline
Constraint                     & A property   that restricts the behavior of a system function                                                                & Bookmarked   section of a document                          \\ \hline
Scenario                        & A description of the interaction of an actor with the system to reach a certain objective                                  & Bookmarked   section of a document                          \\ \hline
OO requirement    & \multicolumn{2}{|l|} {A requirement expressed in a programming language} %                                                     & Bookmarked   section of a document                          
\\ \hline
OO component                  & An abstract   representation of a component of a system or its environment in the form of a   class                      & Class                                                       \\ \hline
OO functional  requirement    & A feature of   a requirements class that realizes a functional requirement                                               & Feature                                                     \\ \hline
OO constraint                   & An assertion   specifying an abstract property of a feature or of a class                                                & One or more   pre- or postconditions or class invariants  \\ \hline
OO scenario                     & An abstract   representation of a scenario in the form of a routine, exercising the   routines of implementation classes & One or more   specification drivers                         \\ \hline
Test           & \multicolumn{2}{|p{14cm}|} {A programming language artifact used to test part of the functionality of a system}%                                                      & Bookmarked   section of a document                          
\\ \hline
Test case                     & A single   testing scenario                                                                                              & Feature                                                     \\ \hline
Test suite                    & A set of   tests covering one or more classes or features                                                                & Class or   cluster                                          \\ \hline
Implementation                        & \multicolumn{2}{|l|} {Code artifacts that are part of the developed system.} % & Bookmarked   section of a document                        }  
\\ \hline
Implementation   component     & A class or   cluster of the implemented system                                                                           & Class                                                       \\ \hline
Implementation   feature       & A feature of   the implemented system                                                                                    & Feature                                                     \\ \hline
\end{tabular}
\caption{Project element types.}
\label{project_element_types}
\end{table*}

\subsection{Relations between project elements}
\label{relations}

The second key element of the model of requirements engineering activity is the set of relations between project elements. Typed relations simplify traceability analysis: to assess the impact of change of a given requirement, selecting links of a certain type will yield the elements that have a certain relation to the requirement. Furthermore, it is possible to introduce formal properties of the relations, which can serve as a basis for relation propagation.

%This section does not attempt to cover all possible types of relations between project elements. Instead, it introduces some basic types and explores formal properties which can be expressed for these relations.

For each relation (except for the reflexive ones) an inverse relation will be defined. Inverse relations are particularly useful for propagating two-way traceability. The summary of project relations and their properties is presented in Table~\ref{Relations_table}.

\begin{table*}[htbp]
\centering
\renewcommand{\arraystretch}{1}
\begin{tabular}{|p{2.55cm}|p{2cm}|p{10.5cm}|}
    \hline
    \textbf{Relation} &
      \textbf{Element  types} &
      \textbf{Definition and formal properties} \\ \hline
    Repeats &
      \begin{tabular}[c]{@{}l@{}}A: ELEM\\ B: ELEM\end{tabular} &
      \begin{tabular}[c]{@{}l@{}}A repeats B if anything which is described by A is   also \\ described by B, and anything which is described by B \\ is also described by A.\\ A repeats B $\implies$ B repeats A\\ A repeats B; B $R_1$ C $\implies$ A $R_1$ C \\ A repeats B; A $R_2$ D $\implies$ B $R_2$ D \\ (where $R_1$, $R_2$ are some relations).\end{tabular} \\ \hline
    Complements &
      \begin{tabular}[c]{@{}l@{}}A: ELEM\\ B: like A\end{tabular} &
      \begin{tabular}[c]{@{}l@{}}A complements B if A and B cooperate towards \\ the   achievement of some higher aim.\\  A complements B $\implies$ B complements A\end{tabular} \\ \hline

    Constrains &
      \begin{tabular}[c]{@{}l@{}}A: CSTR\\ B: FR\end{tabular} &
      \begin{tabular}[c]{@{}l@{}}Requirement A constrains another requirement B if \\  it states a condition that B must satisfy.\\ constrains$^{-1}$ = is\_constrained\_by \\ A constrains B, C constrains B $\implies$ A complements C\end{tabular} \\ \hline
      
    Refines  &
      \begin{tabular}[c]{@{}l@{}}A: ELEM\\ B: ELEM\end{tabular} &
      \begin{tabular}[c]{@{}l@{}}A refines B if anything which is  described by A \\ is also described by B (but some things may be \\ described by B   which are not described by A).\\ (refines; refines) - id $\subseteq$ refines\\ inherits $\subseteq$ refines \\ refines$^{-1}$= generalizes; generalizes$^{-1}$= refines \end{tabular} \\  \hline
    Implements &
      \begin{tabular}[c]{@{}l@{}}A: NLRQ\\ B: IM\end{tabular} &
      \begin{tabular}[c]{@{}l@{}}An implementation feature or implementation \\ component A   implements a requirement B, if it provides \\ the   functionality or constraint   stated in B. \\  implements$^{-1}$ = specifies; specifies$^{-1}$ = implements \\ A refines B $\implies$ A implements B (where:\\ A is implementation feature or component;\\ B is requirement or constraint)\end{tabular} \\ \hline

      \begin{tabular}[c]{@{}l@{}}Contains \\ \end{tabular} &
      \begin{tabular}[c]{@{}l@{}}A: ELEM\\ B: ELEM\end{tabular} &
      \begin{tabular}[c]{@{}l@{}}A contains B if B is a constituent of   A.\\  contains$^{-1}$ = part\_of; part\_of$^{-1}$ = contains \\ (part-of; contains) - id $\subseteq$   complements;\\ (contains; contains) - id $\subseteq$    contains\end{tabular} \\ \hline

    Tests  &
      \begin{tabular}[c]{@{}l@{}}A: TEST\\ B: ELEM\end{tabular} &
      \begin{tabular}[c]{@{}l@{}}A tests B if passing A is required (but possibly not \\  sufficient) for B to be considered implemented correctly\\ tests$^{-1}$ = is\_tested\_by\\ A tests B, C tests B $\implies$ A complements C\end{tabular} \\ \hline    
    Validates  &
      \begin{tabular}[c]{@{}l@{}}A: TEST\\ B: NLRQ\end{tabular} &
      \begin{tabular}[c]{@{}l@{}}A validates B if passing A is required for B to be \\ considered implemented correctly\\validates$^{-1}$ = is\_validated\_by\\ A tests B, B refines C $\implies$ A validates C (where:\\ A is a test artifact; C is a NL requirement)\end{tabular} \\ \hline
    Refers to &
      \begin{tabular}[c]{@{}l@{}}A: ELEM\\ B: NLRQ\end{tabular} &
      \begin{tabular}[c]{@{}l@{}}A refers to B if A refers to B by its name.\\Is\_a\_client $\subseteq$  refers\_to\\ (refers\_to; refers\_to) - id $\subseteq$    refers\_to\end{tabular} \\ \hline

        \end{tabular}
    
    \caption{Relations between project elements and their properties.}
    \label{Relations_table}
\end{table*}

\begin{figure*}[htb!]
    \centering
    \includegraphics[width=0.6\linewidth]{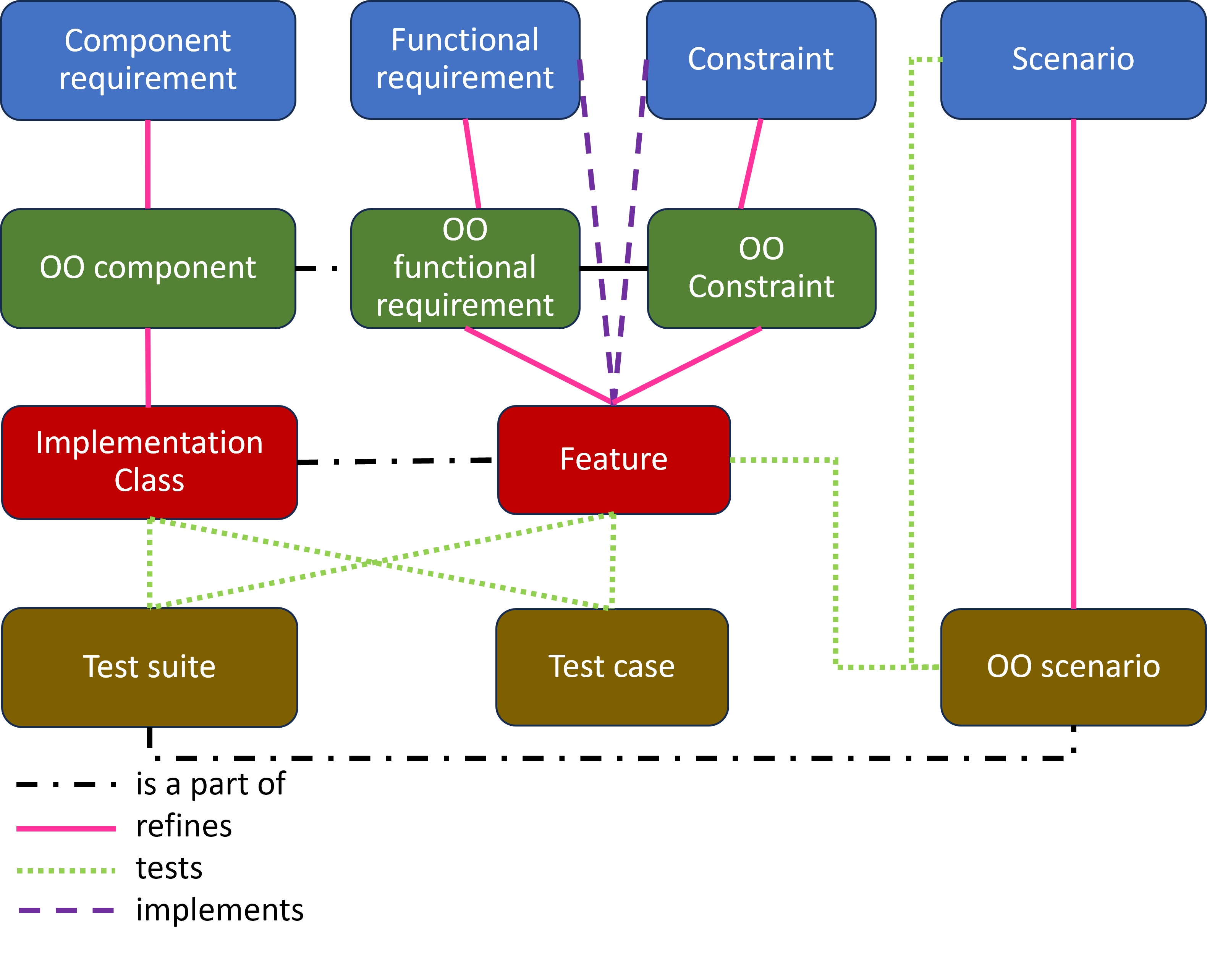}
    \caption{Key project elements and relations.}
    \label{fig:project_elements_graph}
\end{figure*}

Figure~\ref{fig:project_elements_graph} presents project elements and some of the relations between them (some of the relations are not displayed for the sake of clarity).

%Table~\ref{Relations_table} presents the summary of relations' definitions and formal properties. 

\subsection{Propagation of relations}
Requirements traceability is \textit{``the ability to describe
and follow the life of a requirement, in both forward and
backward directions (i.e., from its origins, through its
development and specification, to its subsequent deployment
and use, and through all periods of ongoing refinement and
iteration in any of these phases)''} \cite{gotel1994analysis}. The ability to trace a requirement to implementation and testing artifacts eases the process of validating delivered software against requirements and ensures the adaptability of software products to requirement changes. Manual establishment of traceability links, however, is tedious and time-consuming. Seamless software development may address this issue by providing link propagation so that the links are inferred from the initially created links, project element types, and formal properties of relations between project elements.

To illustrate the idea, let's consider the ``refine'' relation. In seamless software development, a requirement turns into an implementation as the result of a sequence of refinement steps. 
The first step -- linking a natural language requirement with its OO implementation -- is manual, yet such a link can be established in a few clicks, like in most requirements management tools in the field. 
An OO requirement is further linked with the implementation artifact via inheritance: a class implementing a requirement inherits from the requirements class. 
Not necessarily this relation is direct: in fact, there may exist several layers of inheritance. 
However, it is possible to extract the ancestors for implementation artifacts: for each class, it is possible to extract its ancestors and proper ancestors; for each feature, it is possible to identify in which ancestor class it was introduced. 
This way, it is possible to propagate the refinement links based on their formal properties (transitivity). 

When all the ``refine'' links are propagated for a given requirement, the ``implements'' relation can be inferred: if a natural language requirement and an implementation artifact are linked with the ``refines'' relation, they have to be linked with the ``implements'' relation.

%\section{Conclusion}
%This chapter introduced the notion of seamless requirements traceability. It demonstrated that by modeling requirements engineering activity as a set of project elements connected with typed links, it is possible to propagate traceability links based on formal properties of relations. As the same time, if it is possible to propagate traceability links between requirements and related implementation code, it is possible to track changes in requirements to the related code artifacts.

%Seamless requirements traceability, while a theoretical concept, requires  tool support: as it was established before, adequate tool support is one of the key issues in requirements traceability. Current functionality of EiffeStudio, however, does not support typed project elements and relations. The next chapter will present the Traceability tool, which serves as a proof of concept for seamless requirements traceability management.

\section{UOOR traceability tool}
\label{traceability_tool}

One of the key benefits of seamless software development is the ability to facilitate traceability between requirements and other project artifacts, such as implementation and tests. This ability, however, requires adequate tool support. A number of mature tools for traceability links creation and management exist in the market, although they fall short of providing full seamless traceability from requirements to code. 

\begin{comment}
    
To exploit all benefits of seamlessness, the following functionality is required: 
\begin{itemize*}
    \item To link code artifacts (features and classes) with external documents.
    \item To receive notifications of changes in requirements and to trace from requirements to related project artifacts (implementation and tests).
    \item To create typed relations between project elements.
    \item To infer traceability links base on the formal properties of the relations.
\end{itemize*}

\end{comment}
The recently developed UOOR Traceability tool \cite{zakaria} serves as a proof of concept for traceability links management in the UOOR approach. 
The tool is an addition to the EiffelStudio IDE, to which it adds the following mechanisms:
\begin{itemize*}
    \item Assigning types to project elements.
    \item Creating typed links between project elements.
    \item Displaying related links for a given project element.
    \item Displaying notifications of changes in requirements for a given project element.
\end{itemize*}

The Traceability tool is built on top of the Eiffel Information System (EIS), to which it adds four 4 buttons: ``Traceability links'', ``Add link'', ``Delete link'' and ``Track changes''.

\begin{figure*}[t]
  \centering
    \includegraphics[width=\textwidth]{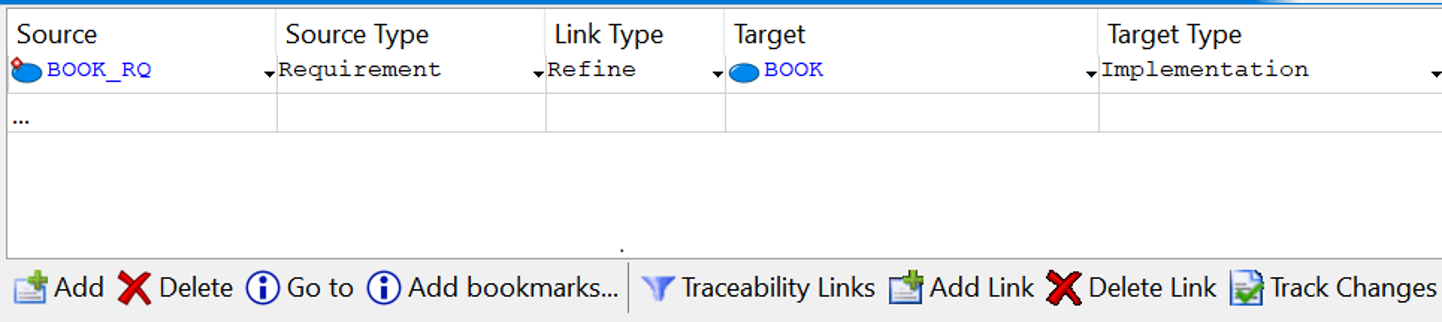}
  \caption{Link management in Traceability tool.}
  \label{fig:link_management}
\end{figure*}

\textbf{Traceability links management.} When adding a new link, a user assigns types to project elements, connected with the link. A link source is a code element (class or a feature). The link target can be a code element or an external document (a bookmarked section of a document). The tool automatically annotates an external document with bookmarks by assigning a bookmark to each paragraph of a text document. Once a traceability link has been created, we can visualize it directly within the code. The tool provides a dedicated interface to view and manage such links (Fig.~\ref{fig:link_management}).

%\mb{add figure with links table}

\begin{figure*}[b]
  \centering
    \includegraphics[width=\textwidth]{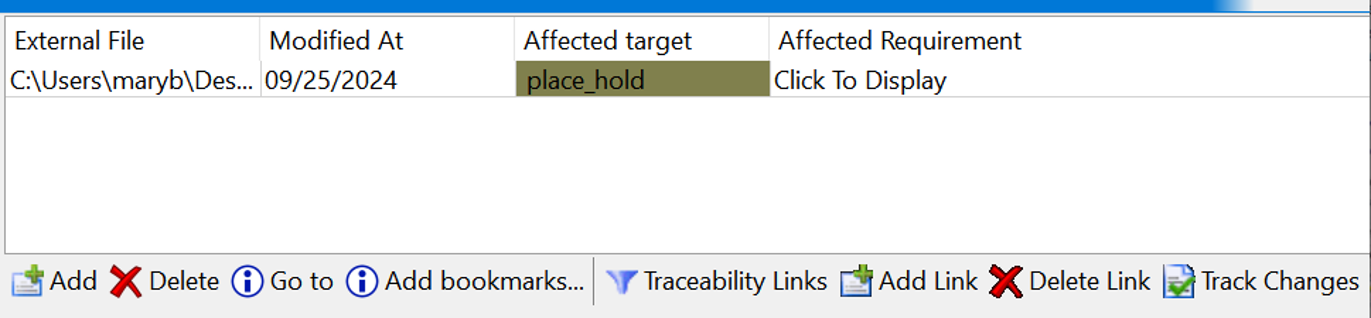}
  \caption{Tracking changes in Traceability tool.}
  \label{fig:track_changes}
\end{figure*}

\textbf{Tracking changes.} The main functionality a project may expect from enforcing requirements traceability is the ability to identify where and how a change in requirements may affect the source code. The Traceability tool addresses this issue by linking requirements in a requirements document with the related code elements, and monitoring changes in text requirements.

The ``Track changes'' button makes it possible to find out changes in requirements and their consequences. If it finds relevant changes, the tool compares the content at each bookmark with the previous version, presenting the outcome in a table showing pairs of requirements and code elements, as illustrated in Fig.~\ref{fig:track_changes}. The table displays the path to the requirements document, modification date, affected project element, and modified requirement. Clicking on ``Click do display'' opens the new version of the modified requirement so that a developer can make relevant adjustments in the code.

\begin{comment}
    
\subsection{Tool limitations and evaluation}
The Traceability tool, developed %as a proof of concept 
for the UOOR approach, has demonstrated that it is feasible:
\begin{itemize*}
    \item To established typed links between software project elements within the IDE.
    \item To track changes in requirements and receive notifications that a requirement, related to a given code element, has been modified.
\end{itemize*}

Being a proof of concept, this tool has a lot of room for improvement, in particular:
\begin{itemize*}
    \item Improving user interface by providing customizable dashboards at different levels of granularity.
    \item Implementing links propagation.
    \item Providing advanced notification system, integrated with communication tools (email, messengers).
\end{itemize*}

\end{comment}

\section{Roborace case study}
\label{Roborace}
Beyond the simple examples used above, an exploratory study has applied the UOOR approach to a practical project, using standard guidelines for empirical software engineering research \cite{casestudy_guidelines}. The example covers some of the requirements of the Roborace system for self-driven race cars. The case study was conducted in collaboration with the autonomous systems group of Constructor Tech (formerly SIT Autonomous) \cite{SIT_autonomous}.

The objective of the study is to explore the following research questions:

%\mb{add citation}
%Add bibliographic reference to https://institute.constructor.org/events/driving-the-future-with-ai

\begin{itemize*}
    \item Is the UOOR approach expressive enough to formulate requirements for a significant project?
    \item Does it facilitate requirements specification?
\end{itemize*}

Roborace is a global championship between autonomous cars. The hardware (the race cars) is the same for all participating teams; each gets access to an autonomous race car called Devbot 2.0 and develops software to drive it in races completely autonomously. Each season sees changes in the goals and rules and the introduction of new conditions.

Use cases helped provide a coarse-grained overview of the race-car functionality; Table~\ref{raceUC}  shows one example, \textit{``Race without obstacles''}, in the Cockburn's style of use-case specification \cite{cockburn}.

\begin{small}
    
\begin{table*}[htb!]
\centering

\begin{tabular}{ |p{0.17\textwidth}|p{0.77\textwidth}| } 

%\centering
%\begin{tabular}

 \hline
 Name & Race\_no\_obstacles \\[4pt]
  \hline
 Scope & System  \\ [4pt]
  \hline
 Level & Business summary  \\ [4pt]
 \hline
 Primary actor & Race car Operator \\[4pt]
 \hline
 Context of use & Race car has to obey an instruction \\ [4pt]
 \hline
 preconditions  & * Race car is on the racetrack grid.\\
                & * Race car is not moving. \\
                & * The global plan (trajectory and velocity profile) minimizing the race time is calculated. \\
                & * The green flag is shown. \\[4pt]
 \hline
 Trigger & The system receives a request from the race car operator to start the race \\[4pt]
 \hline
  Main success scenario & * The system calculates the local plan (path and velocity profile) during the race, trying to follow the global plan as closely as possible. \\
                        & * The race car follows the local plan.\\
                        & * After finishing the required number of laps the race car performs a safe stop.  \\[4pt]
\hline
Success guarantee & The race car has completed the required number of laps and stopped.\\[4pt]
\hline
Extensions
 & A. The red flag is received during the race \\
 & * The race car recalculates a global plan to perform an emergency stop. \\
 & * The race car performs an emergency stop.\\
 & B. The yellow flag is received during the race.\\
 & * The system sets the speed limit according to the received value.\\
 & * The race car finishes the race following the global trajectory and not exceeding the new speed limit.\\
 & C. The difference between the calculated (desired) location and the real (according to the sensors) location is more than a given threshold.\\
 & * The race car recalculates a global plan to perform an emergency stop. \\
 & * The race car performs an emergency stop.\\[4pt]
 \hline
Stakeholders 
& Race car Operator (requests the car to start the race). \\
and interests & Roborace Manager (sets the race goals and policies).\\
& Roborace Operator (shows the green, yellow, and red flags).\\[4pt]
 \hline
%\end{tabular}
\end{tabular}
\caption{A detailed description of the \textit{``Race without obstacles''} use case.}
\label{raceUC}
\end{table*}
\end{small}

\subsection{Modeling components of the system and its environment}
%In autonomous driving domain the system interacts with many environment components. A dedicated effort was made to make all assumptions and constraints explicit. The interviews were combined with domain exploration (by studying related papers and normative documents) to ensure that important environment properties are not overlooked. 

We identified the core components of the system at the start of the project, making it possible to assign functionality to particular modules; the rest of the case study focused on one of them, the planning module. The \e{RACECAR} class (listing~\ref{yellow_flag_constraint}) captures properties applicable to the system as a whole.

\begin{comment}
    
\begin{lstlisting}[caption={Example of a system class.}, captionpos=b, label=racecar_class]

class 
    RACECAR 
feature 
    control_module: CONTROL_MODULE 
    perception_module: PERCEPTION_MODULE 
    planning_module: PLANNING_MODULE 
    localization_and_mapping_module: LOCALIZATION_AND_MAPPING_MODULE
end
\end{lstlisting}
Listing~\ref{racetrack_class} provides an example of an environment class.

\begin{lstlisting}[caption={Example of an environment class.}, captionpos=b, label=racetrack_class]

class 
    RACETRACK 
feature 
    raceline: RACELINE
      -- Optimal raceline for the track 
    map: MAP
      -- Coordinates of the bounding lines 
end

\end{lstlisting}
\end{comment}

The complexity of the domain required splitting the environment into three clusters:
\begin{itemize*}
    \item Environment - elements of the system's environment, with such classes as \e{RACETRACK}, \e{MAP} and \e{OBSTACLE}.
    \item Interfaces - the race car's sensors and actuators, which are external to the developed software, but are part of the cyber-physical system and serve as interfaces between the system and its environment. They include such classes as \e{SENSOR} and \e{LIDAR}.
    \item Auxiliary classes - classes that define abstract concepts of the environment, that are not identified as the environment's elements. They include such classes as \e{LOCATION} and \e{ORIENTATION}.
\end{itemize*}

Inheritance makes it possible to organize the description of these concepts into hierarchies. For example, classes describing sensors, such as \e{LIDAR} and \e{CAMERA}, inherit from \e{SENSOR}, which captures the characteristics applicable to all sensors. In addition, each class describing a particular sensor type captures specific features of that sensor. 

\begin{comment}
    
\begin{lstlisting}[caption={Code of the \e{SENSOR} class}, captionpos=b, label=sensor_class]
deferred class
    SENSOR
feature
    position: LOCATION_3D
      --location in the world coordinates of the scene
    update_rate: REAL
      --sensor update rate
end

\end{lstlisting}  
%\end{figure}

The \e{SENSOR} class has such features as \e{position}, which captures the sensor location on a vehicle, and \e{update_rate} which reflects the frequency of sensor output.  The \e{LIDAR} class inherits from \e{SENSOR} and adds such features as \e{point_cloud} and \e{orientation}.

\begin{lstlisting}[caption={Code of the \e{LIDAR} class.}, captionpos=b, label=lidar_class]

deferred class
    LIDAR
inherit
    SENSOR
feature
    point_cloud: ARRAY2 [LOCATION]
		-- m by n matrix of detected points in lidar coordinate system

    object_points_distance: ARRAY2 [REAL]
		-- m by n matrix of distances to object points

    orientation: ORIENTATION
		-- Lidar orientation in the world coordinates of the scene	
end


\end{lstlisting} 

\end{comment}

General environment constraints, such as the constraint \textit{``If the yellow flag is, up race cars should limit their speed to a dedicated ``safe speed''''} are expressed through the class invariants, as in the following extract from the \e{RACECAR} class (listing~\ref{yellow_flag_constraint}).

\begin{lstlisting}[caption={Implementation of the constraint \textit{``If the yellow flag is up race cars should limit their speed to a dedicated safe speed.''}}, captionpos=b, label=yellow_flag_constraint]
class RACECAR feature
    control_module: CONTROL_MODULE 
    perception_module: PERCEPTION_MODULE 
    planning_module: PLANNING_MODULE 
    localization_and_mapping_module: LOCALIZATION_AND_MAPPING_MODULE

    green_flag_is_up, yellow_flag_is_up, red_flag_is_up: BOOLEAN
    safe_stop_activated: BOOLEAN
    max_speed: REAL	
    current_max_speed: REAL		
        -- Current speed limit
    safe_speed: REAL
        -- Safe speed limit
invariant
    yellow_flag_is_up implies current_max_speed = safe_speed
    green_flag_is_up implies current_max_speed = max_speed
    red_flag_is_up implies safe_stop_activated
end
\end{lstlisting}

Environment assumptions are specified with pre- and postconditions, as the assumption related to the possible sequences of raising flags\footnote{A flag is a signal sent to a race car to indicate a change in racing conditions.} by the Roborace (listing~\ref{roborace_flags}). 

\begin{lstlisting}[caption={Illustration of an environment assunption.}, captionpos=b, label=roborace_flags]
    
class ROBORACE feature
    raise_yellow_flag 
        require green_flag.is_up 
        do 
        ensure 
            yellow_flag.is_up 
            not green_flag.is_up 
            not red_flag.is_up
        end 
    raise_red_flag 
        require green_flag.is_up or yellow_flag.is_up
        do 
        ensure 
            red_flag.is_up 
            not green_flag.is_up 
            not yellow_flag.is_up
        end
    end

\end{lstlisting}

\vspace{-6pt}
\subsection{Producing OO functional specification}
In the Roborace project, each mission focuses on the accomplishment of some scenarios, such as ``Race without obstacles''. Thus the elements of the system's functionality were extracted from scenarios. 

Initially, the system's functions appear simply as the features' names. 
The features are further enriched with contracts, formulating the requirements. Below is an implementation of the requirement ``At every position on a raceline the speed in the velocity profile shall not exceed the race car's maximum speed'' (listing~\ref{calculate_raceline2}).

\begin{comment}
    
(list.~\ref{calculate_raceline}): 

\begin{lstlisting}[caption={Initial implementation of a requirement ``Planning module shall calculate a raceline''.}, captionpos=b, label=calculate_raceline]
class PLANNING_MODULE feature
    calculate_race_trajectory (circuit_map: MAP; vehicle_param: VEHICLE_PARAMETERS; strategy: INTEGER)
      -- Calculate the optimal racing line for a circuit   
        do
        end
end
    
\end{lstlisting}

\end{comment}

\begin{lstlisting}[caption={Implementation of a requirement \textit{``At every position on a raceline the speed in the velocity profile shall not exceed the maximum race car's maximum speed''}.}, captionpos=b, label=calculate_raceline2]
class PLANNING_MODULE feature
    calculate_race_trajectory (circuit_map: MAP; vehicle_param: VEHICLE_PARAMETERS; strategy: INTEGER)
      -- Calculate the optimal racing line for a circuit   
        do
        ensure
            velocity_limit_obeyed: across raceline.velocity_profile as rl all rl.item < car.max_speed end
        end
end
    
\end{lstlisting}

\begin{comment}
    
Contracts also capture abstract properties derived from time-ordering constraints, such as that the global plan must be calculated before calculating the local plan (listing~\ref{local_plan}):

\begin{lstlisting}[caption={Capturing logical constraints.}, captionpos=b, label=local_plan]

class PLANNING_MODULE feature 
    calculate_local_plan: LOCAL_PLAN
      -- Calculate local path from current location 
        require car.global_plan_is_calculated
        do
        ensure car.local_plan_is_calculated
        end
\end{lstlisting}

\end{comment}

\subsection{Producing OO behavioral specification}
%\sectionmark{Roborace: integrating the use cases}
The \textit{``Race without obstacles''} use case, previously expressed in a tabular format, simply becomes a routine \e{race_no_obstacles} in the requirements class \e{ROBORACE_USE_CASES} sketched below (listing~\ref{race_no_obstacles_UC}). Where the use case has alternative flows, the routine  uses conditional expressions.

\begin{lstlisting}[caption={Implementation of the use case \textit{``Race without obstacles.''}}, captionpos=b, label=race_no_obstacles_UC]

race_no_obstacles   
    require
        not car.is_moving 
        car.global_plan_is_calculated
        car.green_flag_is_shown 
        car.is_on_starting_grid
    local local_plan: RACELINE
    do    --Sequence of system actions in use case main flow
        from
        until 
            car.race_is_finished or 
            car.red_flag_is_shown or 
            car.location_error_is_detected 
        loop
            if car.yellow_flag_is_shown then update_speed end
            local_plan := car.planning_module.calculate_local_plan
            car.control_module.move (local_plan.speed, local_plan.orientation)
        end
        if car.red_flag_is_shown or car.location_error_is_detected 
        then emergency_stop else safe_stop end
    ensure
        not car.is_moving 
        car.is_in_normal_mode implies car.race_is_finished
    end
\end{lstlisting}

\vspace{-6pt}
The \e{race_no_obstacles} routine implements the use case by calling the routines \e{update_speed}, \e{safe_stop}, and \e{emergency_stop}, which themselves implement finer-grain use cases. The ``calls'' relation between routines mirrors UML's  \texttt{<<include>>} and \texttt{<<extend>>} relations between use cases. 

The \e{ROBORACE_USE_CASES} class collects such routines reflecting use cases (listing~\ref{roborace_UC}). 

\begin{lstlisting}[caption={Roborace use-case class.}, captionpos=b, label=roborace_UC]
class ROBORACE_USE_CASES feature
    car: RACECAR
    safe_stop 
        require car.is_in_normal_mode
        do car.control_module.safe_stop
        ensure not car_is_moving
        end
    emergency_stop
        require 
            car.red_flag_is_shown or car.location_error_is_detected
        do car.control_module.emergency_stop
        ensure
            not car.is_in_normal_mode
            not car.is_moving
        end
    update_speed
        require car.yellow_flag_is_shown
        do car.update_max_speed (car.yellow_flag_speed)
        ensure car.max_speed = car.yellow_flag_speed
        end			
    race_no_obstacles 
            --implementation is listed above
    -- Other use cases 
end 
\end{lstlisting}

\subsubsection{Relation between use cases and test cases}
\label{use cases to test cases}

\textit{Use case stories} define test cases for use cases \cite{26}. The class
 \e{ROBORACE_USE_CASE_STORIES} inherits from \e{ROBORACE_USE_CASES} class. It includes a collection of routines corresponding to use case stories. 

When a use case takes the form of a routine with contracts, extracting use case stories from such a routine becomes a semi-automated task. 
For example, the \e{emergency_stop} use case accepts two options in its precondition --- (1) when the red flag is shown or (2) when a location error is detected.
These options map to the following use case stories written according to the UOOR approach (listing~\ref{uCstories}):

\begin{lstlisting}[caption={Use case stories extracted from the \textit{``emergency stop''} use case.}, captionpos=b, label=uCstories]
class ROBORACE_USE_CASE_STORIES inherit ROBORACE_USE_CASES feature
    emergency_stop_red_flag_story
        require car.red_flag_is_shown
        do emergency_stop 
        end
    emergency_stop_location_error_story
        require car.location_error_is_detected
        do emergency_stop 
        end
end
\end{lstlisting}

These routines represent the two different paths through the \e{emergency_stop} use case, characterized by their preconditions.
The connection with the parent use case is visible because the stories call the routine encoding the use case.
The two routines must be exercised at least once with test input that meets their preconditions.

A similar analysis makes it possible to extract 5 use case stories from the \textit{``Race without obstacles''} use case:

\begin{itemize*}
    \item 3 for each possible loop exit condition.
    \item 1 corresponding to the true antecedent of the implication in the second postcondition assertion.
    \item 1 corresponding to the true consequent and false antecedent of the said implication. 
\end{itemize*}

The full collection of the extracted use case stories may be found in a publicly available repository \cite{42}.

\subsection{Lessons learned from the case study}
Project stakeholders often fail to state environment-related information explicitly, as they take it for granted. The UOOR methodology can address the issue by serving as a structural basis for elicitation activities.
The approach requires explicitly specifying environmental components and properties such as assumptions, constraints, and invariants.

The \textit{``Race without obstacles''} use case provides a good illustration of the \textbf{difference between contract-based and scenario-based specification}. As a specification, this scenario expresses, among other properties, that the system calculates a local plan and then follows it. 
It states this property in the form of a strict sequence of operations which, however, only covers some of the many possible scenarios.

It does list extensions, but only three of them, and does not reflect the many ways in which they can overlap. For example:
\begin{itemize*}
    \item 

It can happen that the green flag is shown some time after the yellow flag, but the extensions do not even list it.
\item
In the same way, the red flag can be shown after a yellow flag. 

\end{itemize*}

\noindent An attempt to add extensions to cover all possibilities would have no end, as so many events may occur as to create a combinatorial explosion of possible sequencings.

One way out of this dead end would be to use temporal logic \cite{Pnueli}, which provides a finite way to describe a possibly infinite but constrained set of sequences of events or operations. UOOR relies on a different idea: use logical rather than sequential constraints. Sequential constraints become just a special case: we can express that \e{A} must come before \e{B} simply by defining a condition \e{C} as part of both the postcondition of \e{A} and the precondition of \e{B}. But the logic-based specification scheme covers many more possibilities than just this special case. The specification the example just mentioned is presented in listing~\ref{roborace_flags}. \\

The UOOR approach helped to \textbf{structure the entire process of requirements elicitation and analysis}:
\begin{itemize*}
    \item It required identifying and specifying the elements of the environment, separately from elements of the system.
    \item This process revealed the assumptions and constraints which could be overlooked otherwise.
    \item Analysis of use cases revealed more abstract properties, than time-ordering constraints.
    %\item Specifying use case stories with specification drivers defined the test cases for system's behavior.
\end{itemize*}

\subsection{Discussion}
This case study serves as a proof of concept on a significant ongoing project. It is not, however, a systematic empirical validation and as a consequence does not allow drawing firm conclusions such as a guarantee that the UOOR approach will increase productivity or decrease defects. It illustrates instead the observation that object-oriented technology with logical constraints is more general than scenario-based techniques, encompassing them as special cases.

%\subsection{Conclusion}
%    \item Is UOOR approach expressive enough to formulate requirements for a significant project?
%    \item Does the UOOR methodology facilitate requirements specification?
 
%With respect to the research questions, we can make the following conclusions.

The case study did show that in a specific project the UOOR methodology helped a specific group (the authors) structure the process of requirements elicitation (by helping find the questions to be asked from stakeholders) and analysis (by helping to turn specialized scenarios into more general logical specification).

On the Roborace case study we have demonstrated that with the UOOR approach, one can:
\begin{enumerate}
    \item Express the fundamental abstractions in the form of requirements classes.
    \item Express the fundamental constraints in the form of invariants for these classes.
    \item Express typical usage scenarios with specification drivers. (Unlike the previous two, this task does not make any attempt at exhaustiveness, since examples can only cover a fragment of all possibilities; instead, it concentrates on the scenarios of most interest to stakeholders, and those most likely to cause potential issues or bugs.)
    \item As a consistency check, ascertain that the scenarios (item 3) preserve the invariants (item 2).
\end{enumerate}

More generally, the combination of an object-oriented approach to structure the requirements (1), equipped with invariants (2) as well as other forms of contracts (preconditions, postconditions), with use cases to illustrate the requirements through examples of direct interest to stakeholders (3) and shown to preserve the invariants (4) provides a promising method for obtaining correct and practically useful requirements.

\section{UOOR user study}
\label{Experiment}

A study conducted at the University of Toulouse \cite{experiment_paper} evaluates the perception of the UOOR approach and its potential to be adopted in industry. Since software engineers are already equipped with a set of well-known requirements techniques, their willingness to study and adopt a new approach is based on the following factors:
\begin{itemize*}
    \item Is this approach easy to learn? Will it require much time and efforts to study it?
    \item Does this approach provide an added value? Will it help me to improve requirements specifications?
\end{itemize*}

To address these concerns, the study formulates the following research questions: (i) is limited training sufficient for learning OO contract-based requirements? (ii) does learning contract-based OO requirements techniques help to produce better UML specifications? 

\subsection{Study design}
The study was conducted as a part of the ``OO Analysis and Design'' course at the University of Toulouse. This course, like its many counterparts in other universities \cite{20}, introduces UML as requirements modeling language. 

The study had two parts:
\begin{itemize*}
    \item A controlled experiment \cite{91} based on the ``Instantrame\footnote{Instantrame is a social network on smartphones allowing anyone with an account to share photos}'' case study, where students produced requirements applying two different approaches.
    \item A questionnaire, where students reflected on their experience.
\end{itemize*}

In total, 31 students participated in the study: bachelor's students in their third year and master's students in their first year. Course instructors provided a textual description of a case study to students, which they further used to elicit requirements and produce various requirements artifacts. 

The experiment was split into two parts. In the first part (1.5 hours) theory on UOOR requirements was presented to students. They were already familiar with UML and scenario modeling. Further, they had a task to describe two scenarios for each of the two given use cases according to a provided template. In the second part (4.5 hours), students were randomly split into two groups and had to complete two tasks:
\begin{enumerate}
    \item (2-3 hours) Students of Group 1 specified UOOR requirements for the first use case. Students of Group 2 produced a sequence diagram for the first use case.
    \item (2-3 hours) Students of Group 1 worked on a sequence diagram for the second use case. Students of Group 2 specified UOOR requirements for the second use case.
\end{enumerate}

After submitting the results of their work, students filled in an online questionnaire. The questionnaire included two types of questions. Single-choice questions were formulated as statements that participants had to evaluate based on a Likert scale (`Strongly disagree', `Disagree', `Agree', `Strongly agree', `Neutral' choices). We used those questions to collect quantitative feedback on the use of the UOOR approach. The questionnaire also included open questions to collect additional qualitative feedback, such as the perceived advantages of the approach, the difficulties that participants faced using the approach, what are the potential improvements of the UOOR, and how applying UOOR helps to improve UML-based specifications.

\subsection{Study results}
All of the 31 study participants have participated in a survey, yet one of them answered only part of the questions. Although the population size is not large enough to draw definite conclusions, the study provides a preliminary outlook on the usability of the UOOR approach.

%------------------------------------------------------------------
\begin{figure*}[!bhtp]
\includegraphics[scale=1]{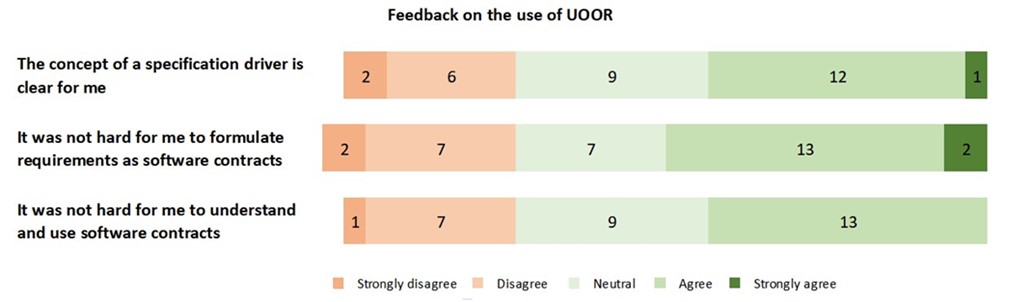}
\caption{Feedback on the use of UOOR.}
\label{UOOR_feedback}
\end{figure*}
%------------------------------------------------------------------

26\% of respondents declared that it was hard to understand and use software contracts for requirements specification, whereas 43\% had a positive experience and 30\% were neutral (Fig.~\ref{UOOR_feedback}). 

%\clearpage

%\newpage

All participants were able to list the advantages of formulating requirements in UOOR. The participants highlighted that the UOOR methodology is easy to understand and follow and that it facilitates producing detailed, specific, readable  requirements which are easy to validate 

\begin{comment}

(Table~\ref{tab:table4}). 

\begin{table}[h!]
    \begin{tabular}{|p{0.95\textwidth}|}
        \hline
        \textbf{What are the reasons to use UOOR?}\\[4pt]
        \hline
        Readability and clarity\\[4pt]
        \hline
        Brings more details at the specification level\\[4pt]
        \hline
        Modularity\\[4pt]
        \hline
        The methodology is easy to understand and follow\\[4pt]
        \hline
        Facilitates identification and analysis of use case scenarios\\[4pt]
        \hline
        Obliges to write preconditions and postconditions\\[4pt]
        \hline
        Well-detailed, very specific\\[4pt]
        \hline
        Easy reuse, which facilitates maintenance\\[4pt]
        \hline
        Better possibility of requirements’ validation\\[4pt]
        \hline
    \end{tabular}
    \caption{Summary of responses to the question “What are the reasons to use UOOR?”}
    \label{tab:table4}
\end{table}

\end{comment}

The particular difficulties stated by the experiment participants, were: not enough familiarity with contracts; not enough examples provided; not enough practice. 

\begin{comment}
    
(Table~\ref{tab:table3}).
 
\begin{table}[h!]
    \begin{tabular}{|p{0.95\textwidth}|}
    \hline
        \textbf{What difficulties have you faced when applying UOOR?}\\[4pt]
        \hline
       Not enough familiarity with Eiffel language and contracts \\[4pt]
        \hline
        It is difficult to formulate pre- and postconditions for some scenarios.\\[4pt]
        \hline
        Need more examples in order to adapt\\[4pt]
        \hline
        Need more practice\\[4pt]
        \hline
    \end{tabular}
    \caption{Summary of responses to the question “What difficulties have you faced when applying UOOR?”}
    \label{tab:table3}
\end{table}

\end{comment}

The questionnaire responses indicate that UOOR helped students to improve their UML specifications in the following ways: “\textit{to think of elements we had not thought of, for example, additional preconditions}”, “\textit{to discover details that need to be added to features}”, “\textit{to define and implement use cases better}”, “\textit{to identify alternative scenarios for the system’s use cases}”, “t\textit{o analyze better the requirements in a global way for a specification}” %(Table~\ref{tab:table5}). 
This is a significant result since people are more likely to adopt an approach which provides immediate benefits, such as improving requirements specifications produced with the currently used approach. 

\begin{comment}
    
\begin{table}[h!]
    \centering
    \begin{tabular}{|p{0.95\textwidth}|}
        \hline
        \textbf{How applying UOOR helped to improve UML specification?}\\[4pt]
        \hline
        Facilitates discovering details that need to be added to features\\[4pt]
        \hline
        Facilitates better analysis of the requirements in a global way\\[4pt]
        \hline
        Understanding of the purpose of the steps to be carried out \\ when writing specifications in UML\\[4pt]
        \hline
        Facilitates identifying alternative scenarios for the system’s use cases\\[4pt]
        \hline
        Facilitates better definition and implementation of use cases\\[4pt]
        \hline
    \end{tabular}
    \caption{Summary of responses to the question “How applying UOOR helped to improve UML specification?”}
    \label{tab:table5}
\end{table}

\end{comment}

The questionnaire also collected feedback from the participants on the usability of the UOOR approach. The responses indicate that a more detailed description of the approach with more examples would improve its usability, as would diagram and tool support. 

\begin{comment}
    
\ref{tab:table6}. 

 %\clearpage

\begin{table}[h!]
    \centering
    \begin{tabular}{|p{0.95\textwidth}|}
        \hline
        \textbf{What could make the UOOR approach more usable?}\\[4pt]
        \hline
        A tool allowing the construction of an architecture according to requirements\\[4pt]
        \hline
        Diagrams to make it more concrete and understandable\\[4pt]
        \hline
        A more detailed description of the approach\\[4pt]
        \hline
        Having more examples\\[4pt]
        \hline
    \end{tabular}
    \caption{Summary of responses to the question ``What could make the UOOR approach more usable?''}
    \label{tab:table6}
\end{table}
\end{comment}

\subsection{Discussion}
Since the experiment was limited to a single course at the University of Toulouse, it is not possible to guarantee that the results transpose to other environment. Another limitation is that the subjects of the experiment were students, rather than software engineering professionals. Note, however, that studies have shown  that there is little difference between the results of software engineering students and professionals for relatively small tasks \cite{host2000using}.

The difficulties reported by the participants, such as not being familiar enough with the Eiffel language and contracts, indicate that the course prerequisites might not have been explicit enough. Since the Eiffel language and Design by Contracts are part of the standard curriculum of the bachelor's software engineering program, we assumed that all study participants are familiar with these topics. Some deviations arise from the presence of guest master students from other programs. 

Several participants claimed that not enough explanations and examples were provided. The present article provides a detailed methodology with two illustrative examples. The approach will also be published in a textbook (``companion book'' \cite{COmpanion}) accompanied by a website and a GitHub repository, which will serve as a forum for discussions of the method.

\section{Related work}
\label{approaches}
A previous article \cite{46} discussed the role of use cases
in requirements and contrasted them with object-oriented requirements. The present paper extends that original discussion to a full-fledged requirements engineering method. 

A number of requirements approaches share at least some of the objectives of UOOR. The field is a very broad one, with hundreds of proposals. We identified 15 well-documented methods which lend themselves to a point-by-point comparison based on the criteria discussed in section~\ref{characteristics}. Table~\ref{tab:related_work} is an overview of the results.

%None of the reviewed approaches satisfies all of the criteria identified in section~\ref{characteristics}.

NL-based requirements \cite{wiegers, 17, RELAX} are requirements formulated in the form of unrestricted \acrshort{nl} text, or \acrshort{nl} text, restricted in a certain way. NL-based requirements are easy to learn and are supported by a wide variety of tools and education materials. Scenarios (use cases \cite{43}, user stories \cite{14}, and use cases 2.0 \cite{25, 26}) are a powerful requirements technique. Still, they cannot serve as a requirements methodology. NL-based requirements, including scenarios, are prone to ambiguity, which can be partially eliminated by constraining the natural language. Requirements traceability relies on manually created traceability links. Requirements are reused by copy-pasting.

Use cases are an important modeling tool in UML \cite{55}. UML makes it possible to treat use cases as objects, subject to specialization and decomposition. UML use cases can have pre- and postconditions. It is possible in UML to associate contracts with individual operations through  natural language or the OCL (Object Constraint Language) notation. SysML \cite{47}, an extended profile of UML, treats requirements as first-class entities, establishing direct links between requirements and other software artifacts (such as tests). \cite{58} illustrates the requirements specification process with SysML, and \cite{5, 52, 57} provide applications of SysML to all phases of software development. SysML does not provide semantics for requirements, although it is possible to associate contracts with individual operations through natural language or the OCL notation. SysML and UML are standardized notations and not methodologies. 

The Restricted Use Case Modeling approach \cite{62} relies on a use case template and a set of restriction rules to reduce the ambiguity of use case specification and facilitate the transition to analysis models, such as UML class diagram and sequence diagram. The aToucan tool automates the generation of UML class, sequence, and activity diagrams \cite{61}. The tool can generate traceability links from the textual use cases to the generated class diagram, but not to the source code. The approach does not advocate extracting abstract properties from use cases and domain knowledge, such as time-ordering and environmental constraints. 

A Use Case Map (UCM) \cite{3,12} depicts several scenarios simultaneously.
UCMs represent use cases as causal sequences of responsibilities, possibly over a set of abstract components. 
In UCMs, pre- and postconditions of use cases, as well as conditions at selection points, can be modeled with formal specification techniques such as ASM or LOTOS.  
UCMs specify properties of operations in relation to scenario sequences, rather than abstract properties of objects and operations. Use case maps do not provide a framework for requirements traceability and reuse.

Object-Oriented Analysis and Design (OOAD) \cite{8} is a unified methodology for use-case-driven analysis and design, supported by UML \cite{55} as a unified notation. OOAD applies OO techniques (class-based decomposition, OO modeling) to the initial requirements, produced at the earlier stages of the development process. OOAD does not provide a framework for requirements traceability and reuse.

The OO-Method \cite{80} combines conventional OO specification techniques \cite{8} with formal specification, relying on the OASIS object-oriented specification language \cite{lopez1995oasis}. 
The Integranova tool supports the specification process with an interactive interface and automatically generates the implementation code. However, this approach does not provide a framework for requirements traceability and reuse.

In goal-oriented requirements engineering  \cite{78}, \cite{60} requirements are obtained through a series of refinements of high-level goals. With the help of the Objectiver tool \cite{44},
requirements can be linked to other artifacts, such as goals, environment agents, or operations. However, traceability links to natural language requirements documents or implementation artifacts are out of the scope of the approach.

In test-driven development (TDD) \cite{66}, a software engineer writes unit tests before implementing the system’s functionality in small iterations. Unit tests can be viewed as the means of capturing requirements: tests serve as a guide to code writing. In behavior-driven development (BDD) \cite{75, 86}, requirements are formulated as user stories, following a specific template. Dedicated tools transform user stories into parameterized unit tests. TDD and BDD rely on scenarios, which are not abstract enough to be \textit{requirements}: if scenarios attempt to cover \textit{all} possible situations, their number explodes, which impedes requirements traceability. BDD and TDD do not provide mechanisms for requirements reusability and static verification.
   
In the ACL/VF framework \cite{6, 64}, use cases capture requirements, which are further formalized as grammars of responsibilities. Another Contract Language (ACL) contracts (pre- and postconditions and invariants) specify constraints, which scenarios' or responsibilities' execution poses on the system's state. In this approach, the requirements model is decoupled from the candidate implementation: a dedicated binding tool maps elements of the requirements model to the elements of candidate implementation. The approach requires a substantial background: familiarity with design by contract, ACL, and formal grammars. The approach is not seamless and does not provide a framework for requirements reuse.

The Multirequirements approach \cite{32} suggests using a single notation (Eiffel programming language) for requirements, design, and implementation. Requirements are formulated in 3 interconnected layers: natural language, software contracts in programming language, and diagrams. The approach does not provide a methodology and a framework for requirements traceability and reuse.

The PEGS approach attempts to provide a definition and taxonomy of requirements. According to this approach, requirements pertain to a Project intended, in a certain Environment, to achieve some Goals by building a System. Thus, requirements specification consists of four books: Project, Environment, Goals, and Systems, which correspond to each of these components \cite{Handbook}. The approach provides principles and techniques of requirements specification, including seamless OO specification, yet does not provide an explicit methodology. 

The SIRCOD approach \cite{galinier2021seamless} provides a pipeline for converting natural language requirements to programming language contracts. The approach relies on the domain specific language, RSML, for automating conversion from natural language to programming language. In the Seamless Object-Oriented Requirements approach (SOOR), requirements are documented as software classes, which makes them verifiable and reusable \cite{39}. Routines of those classes, called specification drivers, take objects to be specified as arguments and express the effect of operations on those objects with pre- and postconditions. The SIRCOD and SOOR approaches focus on translating existing requirements specifications to contracts expressed in a programming language, rather than extracting abstract requirements from scenarios.

The UOOR method relies on the advancements of the SIRCOD and SOOR approaches but focuses more on the approach's usability and requirements traceability management. 

%\begin{landscape}
\begingroup
\setlength\tabcolsep{0.08cm}.

\begin{table*}
\small 
%\resizebox{\textwidth}{!}
\centering
    \begin{tabular}[M]{|p{2.5cm} | p{1.7cm}| p{1.5cm}| p{1.0cm}| p{1.7cm}| p{1.7cm} |p{1.7cm}| p{1.5cm}| p{1.7cm}|}
    \hline
& Methodology & Required \newline background & Tool \newline support & Requirements reusability & Requirements  \newline verifiability & Requirements \newline unambiguity & Traceability \newline support & Seamlessness \\
\hline
NL-based & Yes & Some & Yes & No & No & Partial&No&No \\
UML and SysML & No & Substantial & Yes & No & Partial & Partial & Partial & No \\
Scenarios & No & Some & Yes & No & No & No & No & No \\
RUCM & Yes & Some & Yes & No & No & Partial & Partial & No \\
Use Case Maps & Yes & Substantial & Yes & No & Yes & Yes & No & No \\
OOAD & Yes & Substantial & Yes & No & Partial & Partial & No & No \\
OO-Method & Yes & Some & Yes & No & Yes & Partial & No & No \\
GORE & Yes & Some & Yes & No & No & Yes & Partial & No \\
TDD & Yes & Some & Yes & No & Yes & Yes & Yes & Yes \\ 
BDD & Yes & Some & Yes & No & Yes & Yes & Yes & Yes \\
ACL/VF & Yes & Substantial & No & No & Yes & Yes & Partial & No \\
Multirequirements & No & Some & Partial & No & Yes & Yes & No & Yes \\
SIRCOD & Partial & Some & Yes & No & Yes & Yes & Yes & Yes \\
SOOR & No & Some & Yes & Yes & Yes & Yes & No & Yes \\
PEGS & No & Some & Partial & No & Yes & Yes & No & Yes \\
\hline
    \end{tabular}
    \caption{Summary of related work.}
    \label{tab:related_work}
\end{table*}
\endgroup
%\end{landscape}

\section{Conclusion and perspectives}
\label{contributions}

In this article, we explored the practical application of seamless requirements by addressing the following research questions:
\begin{itemize*}
    \item What should be the process of seamless software development from requirements to code?
    \item What tool support is required to ensure traceability between requirements and other project artifacts?
\end{itemize*}

In answer to these questions, we presented the Unified Object-Oriented approach to Requirements (UOOR) and the supporting Traceability tool. 
%It demonstrated that the approach is affordable and practically usable and that it facilitates producing unambiguous, verifiable, reusable and traceable requirements. 

The UOOR approach is: 
\begin{itemize*}
    \item \textbf{Explicit: }The approach provides a methodology and illustrative examples. The article serves as a guide for the process of producing requirements. The ongoing work on the book illustrating the approach \cite{COmpanion} and online portal with the book's supporting materials may serve as a discussion forum for the adopters of the approach.
    \item \textbf{Lightweight:} The approach does not require the knowledge of formal models or mathematical notations. It does require familiarity with a programming language and Design by Contract. The clarity of the Eiffel language, chosen as the requirements notation, significantly lowers this barrier: in order to be able to formulate contracts in Eiffel, a requirements engineer must learn only a few language constructs. The ability to rely on the mature IDE, EiffelStudio, makes the task even easier. Since OO requirements are compilable code elements, the IDE provides error codes if OO requirements are formulated with errors.      
    \item \textbf{Tool-supported:} The approach relies on the general-purpose IDE (EiffelStudio for the Eiffel language) and integrated Traceability tool. EiffelStudio provides facilities for static and dynamic verification of the developed system against the requirements, enables traceability links creation and management, and provides syntactic checks for OO requirements. The Traceability tool provides more advanced functionality for traceability links management: it allows assigning types to project elements, creating typed traceability links, and tracking changes from requirements to the related project elements. 
    \item \textbf{Seamless:} The UOOR approach provides a methodology for producing UOOR requirements rather than translating requirements documents to a programming language. Thus, UOOR relies on a uniform process (based on refinement) and a uniform notation (Eiffel language). 
\end{itemize*}

Requirements formulated with the UOOR approach are:
\begin{itemize*}
    \item \textbf{Reusable.} Based on the capabilities provided by the object-oriented technology, requirements in UOOR can be reused not by copy-pasting natural language texts but as libraries of domain-specific component specifications in the form of contracted deferred classes. Being implementation-independent, such specifications allow for different implementations.
    \item \textbf{Verifiable.} Since OO requirements are code elements, an IDE provides basic consistency checks at compile time. When contract checking is enabled at runtime, the IDE monitors contracts violations. Since every contract can have a unique tag, a developer can trace an exception to the violated contract. 
Contract specifications serve as oracles for dynamic testing. Moreover, scenarios implemented as specification drivers serve as tests when actual arguments are passed. 
A static verifier, such as Autoproof \cite{7}, can be used to ensure static verification of the system's functional correctness.
    \item \textbf{Unambiguous.} Ambiguity, innate to natural language texts, can be eliminated by introducing formal notation. In UOOR, contracts serve as the notation for requirements, which are understandable due to the readability and clarity of the Eiffel language; simultaneously, they remove the ambiguity inherent to natural language texts. 
    \item \textbf{Traceable.} The approach introduces the notion of seamless requirements traceability, which enables propagating traceability links based on formal properties of relations between project elements. A dedicated Traceability tool provides facilities for creating and managing traceability links and tracking changes from requirements to related code elements.
\end{itemize*}

\subsection*{Limitations}\label{limitations}
The approach has been so far used in connection with Eiffel, relying on the language's expressiveness, its native support of contracts, and the rich functionality of the supporting tool machinery: EiffelStudio, Autoproof, AutoTest. Applying UOOR in a different OO environment is possible but will lose some of the benefits of an integrated approach. The issue, however, is not all-or-nothing, since Eiffel and the tools have open architectures supporting the inclusion of various technology elements. An intermediate approach is possible in which a core system is built using the Eiffel-integrated approach, which can be integrated with numerous elements in other languages and used in conjunction with other tools. It is also possible to use Eiffel as an intermediate requirements, design and prototyping language and then move to another implementation language, although this approach obviously damages seamlessness. 

%The article explored the application of the approach to the greenfield (i.e., developed from scratch) projects. Its application to the brownfield projects may be tedious in case of poor documentation practices. 

The UOOR approach currently focuses on functional requirements. Its application to the Roborace case study shows that it covers a wide spectrum of requirements; including a systematic approach to non-functional requirements is one of the future work directions discussed next.

\subsection*{Perspectives}\label{perspectives}

Forthcoming steps in the development of the UOOR approach include applying and evaluating the approach on a larger scale, such as industrial projects and related user studies. The ongoing work on the ``Companion'' book illustrating the approach \cite{COmpanion} and a supporting website contributes to creating a community of the users and recruiting projects for  evaluation.  

Another direction of future work is the coverage of non-functional requirements. How can non-functional requirements be treated using the UOOR approach? Recent work on use-case modeling of crosscutting concerns \cite{yue2016practical} can be a starting point in such research. 

The work on seamless requirements traceability can be continued in the following directions:
\begin{itemize*}
    \item Further exploration of relation propagation rules.
    \item Extension of this modeling work to a full theory of relations in requirements engineering, with a set of axioms and a number of resulting theorems.
    \item Practical evaluation of seamless traceability on a significant project.
    \item Tool support.
\end{itemize*}

A recent development is  the release of a Traceability tool \cite{zakaria}, which enables creating typed links between project elements and tracking changes from requirements to the related project artifacts. At the time of writing the Traceability tool is in the process of being integrated with the next (mid-2025) version of EiffelStudio. It is itself subject to a number of planned enhancements, in particular a full implementation of link propagation, which will enable seamless traceability between software project artifacts.

Equipped with a traceability management and link propagation tool, UOOR is an effective and useful requirements analysis method for projects which, while not necessarily ready to adopt a fully formal requirements approach, need a rigorous, methodologically sound approach answering the major issues of requirements engineering and focused on producing requirements that are best poised to play their part in the search for very-high-quality software systems. 
%\mb{add closing sentence here}
%\se{TODO}

%%%% Bibliography before appendices
%%%% After the initial round, you need to fix it by hand depending on where you generate your .bib file from
%%%% Mendeley often messes with italics and accents for instance

%\bibliographystyle{abbrv}
%\printbibliography

\bibliography{all}
% Appendices
%\input{chapters/appendix}

\end{document}